\documentclass{aa}  
\usepackage{mathtools}
\usepackage{natbib}
\usepackage{multirow, bigdelim}
\usepackage{nicematrix}
\usepackage{siunitx}
\usepackage{comment}
\usepackage{booktabs}
\usepackage{graphicx}
\usepackage{menukeys}
\usepackage{algorithm}
\usepackage{ulem}
\usepackage[noend]{algpseudocode}
\usepackage[colorlinks=true, allcolors=blue]{hyperref}
\usepackage{txfonts}
 
\newcommand{\cthp}{C$_2$H$^{+}$}

\usepackage{cuted}
\makeatletter
\renewcommand*\aa@pageof{, page \thepage{} of \pageref*{LastPage}}
\makeatother
\usepackage{lineno}

\newcommand{\felix}{HFML-FELIX, Toernooiveld 7, 6525 ED Nijmegen, The Netherlands}
\newcommand{\imm}{Institute for Molecules and Materials, Radboud University, Heyendaalseweg 135, 6525 AJ Nijmegen, The Netherlands}

\begin{document} 
   \title{First detection of C$_2$H$^+$ in the interstellar medium}

   \author{ 
          Arshia M. Jacob\inst{1,2,3}
          \and Karl M. Menten\inst{2}\thanks{This article is dedicated to the memory of Prof. Dr. Karl M. Menten, who passed away unexpectedly, late 2024. Karl was a driving force behind this research, inspiring the team to push the limits of the APEX telescope and pursue the first detection of \cthp\ and many other molecules in the ISM. His vision, curiosity, and insistence that we can always do better brought this work to fruition. He was an outstanding scientist, an exceptional mentor, and an even greater human being. We are deeply grateful for his infectious enthusiasm and for all that he has taught us.}
          \and
          Sandra Br\"{u}nken\inst{4,5} \and 
          Arnaud Belloche\inst{2} \and Friedrich Wyrowski\inst{2} \and Weslley G.~D.~P.~Silva\inst{1,3} \and Oskar Asvany\inst{1,3} \and
          Sarwar Khan\inst{1,2,3} \and Slawa Kabanovic\inst{1,6} \and Kim Steenbakkers\inst{4, 5} \and Gerrit C. Groenenboom\inst{7} \and Britta Redlich\inst{4,5,8}  \and 
   Stephan Schlemmer\inst{1,3}
          }

   \institute{
       I. Physikalisches Institut, Universit\"{a}t zu K\"{o}ln, Z\"{u}lpicher Str. 77, 50937 K\"{o}ln, Germany \email{ajacob@ph1.uni-koeln.de}
       \and
   Max-Planck-Institut f\"{u}r Radioastronomie, Auf dem H\"{u}gel 69, 53121 Bonn, Germany 
   \email{ajacob@mpifr-bonn.mpg.de}
   \and
   Cluster of Excellence “Our Dynamic Universe” (DYNAVERSE)
   \and
   \felix
   \and
   \imm
   \and 
  Zentrum für Astronomie der Universität Heidelberg, Astronomisches Rechen-Institut, Mönchhofstraße 12-14, 69120 Heidelberg, Germany
   \and 
   Theoretical Chemistry, Institute for Molecular and Materials, Heyendaalseweg 135, 6525 AJ Nijmegen, the Netherlands
   \and 
Photon Science Division, Deutsches-Elektron-Synchrotron DESY, Notkestr. 85, 22607, Hamburg, Germany 
 }

   \date{Received 09 July 2026; accepted 05 August 2026}
  \titlerunning{First detection of C$_2$H$^+$ in the interstellar medium}
   \authorrunning{A. M. Jacob et al.}
 \abstract{Despite the detection of nearly 350 molecules in the interstellar medium, almost half of which are carbon chains, the pathways that build molecular complexity remain poorly understood. Observed abundances of carbon-chain and aromatic species are difficult to reconcile with existing top-down or bottom-up formation scenarios, due in part to limited observational constraints and incomplete theoretical understanding. In particular, small intermediary ions, key drivers of ion–molecule reactions capable of seeding larger hydrocarbons and aromatic rings, could provide critical support for the bottom-up formation scenario. Constraining the abundance and chemistry of these ions is therefore essential to test whether bottom-up growth can operate efficiently under interstellar conditions. Here, we report the first detection of the small hydrocarbon cation ethynylium, \cthp, toward the Orion Bar, based on observations with the APEX 12~m sub-mm telescope of its 
 lowest-lying $J = 3 $--$2$ rotational transition near 211~GHz, which exhibits a unique spectroscopic fingerprint through resolved $\Lambda$-doubling and hyperfine splitting components, as recently measured in the laboratory. We estimate \cthp\ column densities between $0.3 \times 10^{11}$~cm$^{-2}$ and $2.2 \times 10^{11}$~cm$^{-2}$ for excitation temperatures of 14--138~K, corresponding to abundances of at most a few $\times 10^{-12}$ relative to the total hydrogen column. Meudon photodissociation region (PDR) models successfully reproduce these values, placing \cthp\ formation at the outer edges of PDR fronts. Our results link \cthp\ production to CH$^+$ and CH$_3^+$ within a network of ion–molecule reactions driven by vibrationally excited H$_2$, a scenario now further supported by recent detections of these species in PDRs like the Orion Bar with JWST observations. The importance of \cthp\ lies in its role as a key intermediate: it produces C$_2$H$_2^+$ and subsequently C$_2$H$_3^+$, effectively channelling small C$_2$ building blocks toward larger hydrocarbons and facilitating bottom-up growth at the PDR surface. Targeted searches for \cthp\ in other regions promise to provide a potentially decisive probe of ion-driven bottom-up chemistry in the ISM.} 
 \keywords{ISM: molecules -- ISM: abundances -- ISM: clouds -- astrochemistry}

   \maketitle
    \nolinenumbers
%

\section{Introduction} \label{sec:intro}

Although the harsh conditions of the interstellar medium (ISM) create a hostile environment for the formation, survival, and growth of molecules and chemical complexity, a remarkable diversity of species has nevertheless been detected across a range of astronomical environments. These span from simple diatomic molecules to large carbonaceous structures such as fullerenes containing up to 70 carbon atoms \citep{Cami2010, Sellgren2010}. Notably, numerous small hydrocarbon species have been identified in photodissociation regions (PDRs), where they are often found in unexpectedly high abundances \citep{Pety2005,Guzman2015}. These regions, composed primarily of neutral gas and dust, are governed by physical and chemical processes driven by far-ultraviolet (FUV) photons with energies between 6~eV and $13.6~$eV \citep{Sternberg1995}. While most of the FUV radiation is absorbed by dust grains and large carbon molecules, such as polycyclic aromatic hydrocarbons (PAHs), resulting in dust heating, a fraction of this radiation also heats interstellar gas via the photoelectric effect. These photons not only dominate the heating but also drive photodissociation and photoionisation processes, enhancing the gas-phase formation of reactive interstellar radicals and ions, and underpinning the remarkable chemical diversity observed.

Today, despite the detection of nearly 350\footnote{See, \href{https://cdms.astro.uni-koeln.de/classic/molecules}{https://cdms.astro.uni-koeln.de/classic/molecules}} molecules in the ISM over the past 90 years \citep[see][for a census]{Endres2016, McGuire2022}, interpreting the abundances of chemical species--especially that of several hydrocarbons, enhanced by FUV radiation--remains challenging when relying solely on gas-phase chemical models built from bottom-up gas-phase chemistry \citep{Pety2005, leGal2017}. 
\cite{Pety2005} demonstrated that steep density gradients at the edges of PDRs are necessary to reproduce the observed spatial offset between hydrocarbon and H$_2$ emission peaks. However, even after incorporating these gradients, models still failed to match the observed hydrocarbon abundances. To address this discrepancy, these authors proposed the need for additional formation pathways beyond those of standard gas-phase chemistry. Specifically, under intense UV irradiation, the fragmentation of large PAHs and small carbonaceous grains could release small carbon clusters and molecules into the ISM in significant quantities, known as top-down chemistry. Subsequent results from both astronomical observations \citep[including for example,][]{Pilleri2013, Guzman2015} as well as laboratory experiments \citep{Zhen2014, Rap2023, Rap2024} further support this conjecture, reinforcing the importance of incorporating top-down processes in chemical models of PDRs. In contrast, chemical models by \cite{Cuadrado2015} have successfully reproduced the observed high abundances of hydrocarbons in strongly irradiated environments such as the Orion Bar without invoking PAH photodestruction, suggesting that multiple formation pathways may be at play. This difficulty reflects the complex nature of interstellar molecules and their formation, which can proceed through two distinct pathways: bottom-up processes, where smaller molecules build up into larger ones, initiated by rapid radical-neutral and ion-neutral reactions in the gas phase \citep{Murga2020}, or top-down processes, such as the fragmentation of PAHs \citep{leGal2017}, as mentioned above. Therefore, a complete analysis of astrochemical abundances necessitates the joint treatment of both gas-phase and grain-surface chemistry. While exploring these varied formation pathways is essential, explicitly determining the relative contributions of bottom-up versus top-down processes, particularly in the case of small hydrocarbons (the precursors of interstellar complex organic molecules (iCOMs)\footnote{Organic molecules comprising six or more atoms detected in the ISM are collectively termed as interstellar complex organic molecules \citep{Herbst2009}.}), remains challenging \citep{Herbst2021}. This difficulty is compounded by the potential incompleteness of chemical networks, uncertainties in photodissociation rates, and the complexities associated with modelling the fragmentation and reformation of PAHs. 

Given the limited understanding of top-down mechanisms, gaining insights into key gas-phase progenitors is imperative for better constraining astrochemical models. Amongst these astrochemical progenitors, molecular ions are important intermediates in the early stages of gas-phase astrochemical networks, owing to their rapid reactions with neutral species and recombination with electrons. Nevertheless, despite their astrochemical significance, many fundamental ions remain elusive in the ISM \citep{Snow2008, McGuire2020}. \\

In recent years, advancements in receiver technology, and significant progress in the laboratory spectroscopy of molecular ions have helped overcome these challenges. A notable example is the detection of the methyl cation, CH$_3^+$, a cornerstone in interstellar carbon chemistry, in emission toward the protoplanetary disk or proplyd, d203$-$506. This detection was achieved via its rovibrational bands at 7~$\mu$m using the James Webb Space Telescope (JWST) by \citet{Berne2023} and was made possible through the telescope's unprecedented sensitivity at mid-infrared wavelengths. Complementary theoretical and experimental analyses of CH$_3^+$'s rovibrational structure enabled the spectroscopic assignment of its observed features and the derivation of physical quantities of astronomical interest, such as excitation temperature \citep{Changala2023, Salomon2026}. More recently, CH$_3^+$ has also been detected in the disk of T-Tauri star TW-Hya \citep{Henning2024} as well as in the Orion Bar PDR \citep{Zannese2025}. These JWST observations, which were able to resolve H$_2$ emission, also suggest that the formation of CH$_3^+$ may be driven by chemical pumping mechanisms induced by FUV-excited H$_2$, particularly in the warm and highly UV irradiated gas layers of PDRs \citep{Goicoechea2025}. 

In addition to CH$_3^+$, ion-molecule reactions
with cations like C$_3$H$^+$ and C$_2$H$_2^+$ also form important gas-phase channels for the formation of hydrocarbons \citep{Schiff1979} and iCOMs \citep{Herbst2017}. The former, C$_3$H$^+$, was first identified by \citet{Pety2012} toward the Horsehead nebula and only later unambiguously confirmed using advances in cryogenic ion trap spectroscopy techniques by \citet{Bruenken2014}. Very recently, the advent of even more sensitive methods such as Leak-Out Spectroscopy \citep[LOS;][]{Schmid2022, Asvany2023} has enabled the measurement of high-resolution spectra of astronomically relevant molecular ions that had previously evaded laboratory detection, thereby providing highly accurate spectroscopic data for their first astronomical searches and successful detection, as exemplified by the recent detection of H$_2$CCCH$^+$ toward the Taurus molecular cloud using the Yebes 40~m telescope\footnote{Utilising data collected as part of the Q-band Ultrasensitive Inspection Journey to the Obscure TMC-1 Environment (QUIJOTE) survey \citep{Cernicharo2021}.} \citep{Silva2023, Silva2024}. These discoveries further highlight the interdisciplinary nature of astrochemical studies, bridging observational, theoretical, and experimental approaches. 

\begin{figure}
    \centering
    \includegraphics[width=0.47\textwidth]{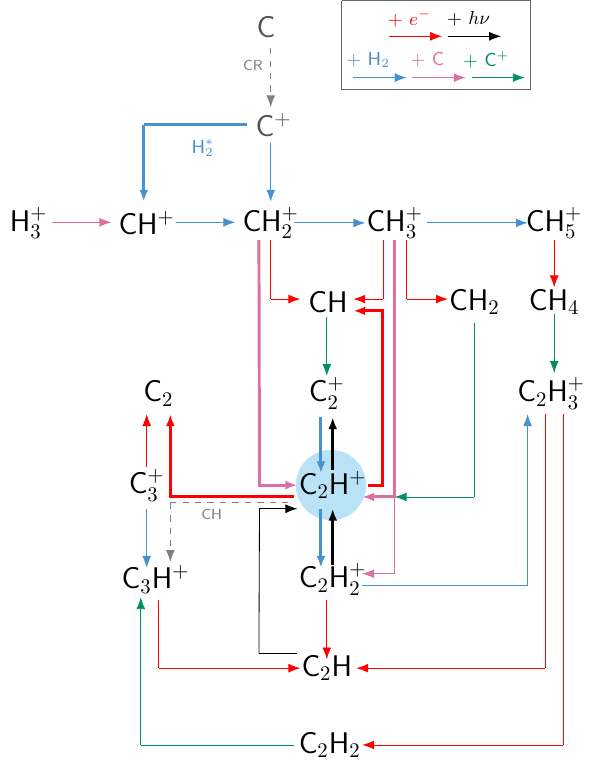}
    \caption{Chemical network displaying the relevant gas-phase ion-neutral and neutral-neutral reactions in the early stages of interstellar carbon chemistry. The dominant formation and destruction pathways of \cthp\ are marked by broader arrows. Unless otherwise labelled, the arrows indicate reactions with electrons, photons ($h\nu$), H$_2$, C, and C$^+$ in red, black, blue, pink, and green, respectively. Adapted from Figure~1 of \citet{Taniguchi2024}.}
    \label{fig:chemical_network}
\end{figure}

As mentioned above, the acetylene ion, C$_2$H$_2^+$, is another key driver of ion-neutral reactions and organic chemistry in the ISM. Lacking a permanent dipole moment and therefore exhibiting no rotational transitions, the astronomical confirmation of this ion requires infrared (IR) observations of its low-lying IR active $\nu_5$ bending vibration  \citep{Asvany2005,Steenbakkers2024} or antisymmetric $\nu_3$ C--H stretching vibration bands  \citep{Jagod1992,Schlemmer2024} at 14 $\mu$m and 3 $\mu$m, respectively, which have not yet been detected in astronomical sources. However, the detection of its chemical precursor, the ethynyl radical cation or ethynylium, \cthp\ (see Fig.~\ref{fig:chemical_network}), could provide valuable constraints on the early stages of hydrocarbon chemistry. Triggered by recent spectroscopic measurements of its rotational spectrum by \citet{Steenbakkers2025}, this paper reports the first detection of \cthp\ toward the CO$^+$ emission peak in the Orion Bar. Sections~\ref{sec:spectroscopy} and \ref{sec:observations} briefly describe the spectroscopic features of \cthp\ and the observations carried out, respectively, with the results presented in Sect.~\ref{sec:results}. The chemical model used in the analysis and the subsequently derived constraints on PDR chemistry are discussed in Sect.~\ref{sec:discussion}, with the main findings summarised in Sect.~\ref{sec:conclusions}. \\

\section{Spectroscopy} 
\label{sec:spectroscopy}

Measuring laboratory spectra of highly reactive ions like ethynylium, \cthp, has been a great experimental challenge until recently. With the development of the LOS \citep{Schmid2022} method in cryogenic ion traps, recording such spectra has reached a new level of quality. \citet{Steenbakkers2025} employed LOS in the cryogenic ion trap setup COLTRAP \citep{Asvany2010,Asvany2014} to investigate the ro-vibrational spectrum of the C-H stretching vibration of \cthp\ in the 3~$\mu$m wavelength range, and subsequently recorded its five lowest rotational transitions using a double-resonance LOS scheme \citep{Asvany2023}. Details of the analysis of those spectra can be found in \citet{Steenbakkers2025}.

\cthp\ is a linear molecule with a $X^{3}\Pi$ ground electronic state. Due to its open shell nature, its rotational spectrum, which we consider in this work, does not simply exhibit the typical harmonic progression of rotational frequencies of a linear rotor where the lines appear with a constant spacing of twice the rotational constant, $B$. 
Instead, the energy term diagram is governed by the spin-orbit coupling 
created by the interaction of the total spin of the two unpaired electrons ($S=1$) and the orbital angular momentum, $L=1$. Based on this fact, the rotational spectrum will exhibit substructure due to coupling effects of the electronic angular momentum ($\boldsymbol{L}$), the electron spin angular momentum ($\boldsymbol{S}$), the rotation of the molecule (angular momentum $\boldsymbol{R}$) and even by the hyperfine interactions of the resulting angular momentum ($\boldsymbol{J}= \boldsymbol{R}+ \boldsymbol{L}+ \boldsymbol{S}$) with the nuclear spin angular momentum ($I=1/2$) of the proton in \cthp, leading to the net angular momentum ($\boldsymbol{F}= \boldsymbol{J}+ \boldsymbol{I}$). It requires involved spectroscopic studies in the laboratory to unfold all these interactions, but at the same time, the unique substructure in the rotational spectrum acts like a fingerprint to detect and identify a molecule even from only one rotational transition, as will be laid out below for our target molecule C$_2$H$^+$.

  Here we discuss only the rotational spectrum because the unique features of the ground state rotational transitions will be used to identify them in the astronomical spectrum. This process is aided by a hierarchy of energies which leads to an ordering of states imprinted in the energy term diagram of a molecule. For the open shell \cthp, the largest energy spacing arises from the fine-structure states $\Omega =0,1,2$ which are separated by about 15~cm$^{-1}$, with $\Omega=2$ being the lowest in energy and thus with the highest population at low temperatures.
  The next largest energy spacing arises from molecular rotation and indeed a coarse separation of 2B ($\sim$ \!2.7~cm$^{-1}$ $\sim$ \!80~GHz) is found in the laboratory spectra when changing the rotational state by $\Delta J=\pm1$. 
  For $\Omega=2$, the lowest value of $J$ is also 2 and thus the lowest 
  rotational transition is $J=3\rightarrow 2$ leading to an energy difference of about 6B, 
  expected at a frequency of about 240~GHz. The six lines belonging to the $J=3\rightarrow2$ transition, as listed in Table~\ref{tab:spec_properties}, appear at  significantly lower frequencies as also observed in the laboratory spectra due to the open shell nature of C$_2$H$^+$  \citep{Steenbakkers2025}. Thus, the coarse position of these lines, $\sim\!211.5$~GHz, is one characteristic of the rotational spectrum of C$_2$H$^+$. The six observed fine-structure lines belonging to the one rotational transition fall into two groups separated by about 200~MHz, due to the two orientations of the orbital angular momentum ($L=1$) with respect to the molecular axis, $\Lambda=\pm1$.

  As a result, each rotational level $J$ splits into two levels with opposite parity, the so-called  $\Lambda$-doubling, labelled here as (e) and (f). For $\Delta J=\pm1$, a given transition's parity changes by the change of $J$. Therefore, for the transitions listed in  Table~\ref{tab:spec_properties} the rotationless parity (e) or (f) does not change. The coarse separation ($\sim$200 MHz) of these groups of lines is another characteristic of the spectrum of \cthp\ and can be nicely seen in the full spectrum displayed in Appendix~\ref{appendix:APEX_full-spec}. At last, the relative orientation of $\boldsymbol{J}$ and the nuclear spin of the proton $\boldsymbol{I}$ leads to the hyperfine structure (HFS) of the resulting total angular momentum states $F=J\pm1/2$, leading to the respective transitions $F' \rightarrow F''$ as specified in Table~\ref{tab:spec_properties}. The corresponding HFS of the three (e) and (f) transitions spreads the three lines each over a range of about 30~MHz which is the third  characteristic of the spectrum. 

A complete account of the experimental laboratory data, their global spectroscopic analysis, 
and the derived spectroscopic parameters based on an effective Hamiltonian 
approach using the PGOPHER software \citep{Western2017} is given in \cite{Steenbakkers2025}. The global fit incorporated five high-resolution rotational transitions (with an accuracy ${\sim\!80}$~kHz), along with additional ro-vibrational transitions (with an accuracy ${\sim\!30}$~MHz), 
connecting the ground state to two vibrationally excited states, 
providing accurate spectroscopic information up to $J=7$ in 
the $\Omega=2$ state. Due to the lower resolution of the ro-vibrational spectra, 
predicted rotational lines in other $\Omega$ states ($\Omega=0,1$), 
as well as potential transitions between $\Omega$ states, 
are significantly more uncertain compared to the lines observed 
in this work, particularly with respect to their HFS. 

Einstein coefficients, $A_{ul}$, and partition functions, $Q$, have 
been calculated using the PGOPHER program. 
A permanent dipole moment of 1.06~D was calculated 
at the MRCI/ANO1 and HF/cc-pVQZ level of theory, but needs to be treated with caution. 
For the partition function calculations, the energy origin was set to the lowest HFS state, 
and levels with $J<30$ were included, 
providing converged values for $Q$ up to 300~K. 
The effect of populating higher $\Omega$ states is more complex to evaluate. 
However, this aspect should be reasonably well accounted for 
as the dominant contributions from spin–orbit, spin–rotation, 
and $\Lambda$-doubling terms are well constrained by the ro-vibrational data. 
In addition, the partition function only includes contributions from 
the vibrational ground state, neglecting excitations to any of \cthp 's many, 
also low-lying vibrational states \citep{Steenbakkers_vibronic}, 
which could be populated at temperatures above 100~K, 
but should not have much influence at the lower temperatures considered here. 
These uncertainties may nonetheless impact calculated line intensities and, 
by extension, abundance estimates. 
 Despite these limitations, the analysis of the laboratory spectrum 
 by \citet{Steenbakkers2025}
 should give a reasonable account to model observed frequencies, intensities and abundances
 for the first astronomical detection of \cthp.

\begin{table*}
{\small 
\centering
\caption{Frequencies, spectroscopic parameters, and derived line-fitting parameters for the $J= 3$--\,2 ($\Omega=2$) of \cthp.}
    \begin{tabular}{llll  lll  c c c c}
    
    \hline\hline
   \multicolumn{4}{c}{Transition} & \multicolumn{1}{c}{$\nu$} & $g_u$ & \multicolumn{1}{c}{$A_{\rm ul}$ \ [s$^{-1}$]} &  $\upsilon_{\rm LSR}$ & $\Delta \upsilon$ & $T_{\rm MB}$ & $\int T_{\rm MB} {\rm d}\upsilon $\\

   $F^\prime$ & $p^\prime$ &  $F^{\prime\prime}$ & $p^{\prime\prime}$ & \multicolumn{1}{c}{[MHz]} & & $\times10^{-6}$ &  [km~s$^{-1}$] & [km~s$^{-1}$] & [mK] & [mK~km~s$^{-1}$] \\
   
\hline
  2.5&  f    & 2.5 &  f  & 211417.82(8)   &  6 & 0.200  & -- & -- & --& --\\ 
  3.5&  f    & 2.5 &  f  & 211435.49(8)    & 8 & 3.001  & 10.23(0.40) & 3.76(0.21) & 3.75 &  15.28(1.54)\\ 
  2.5&  f    & 1.5 &  f  & 211447.14(8)    & 6 &  2.801  & $-$6.57(0.41)  & 3.76(0.21) & 2.62 & 10.70(2.20)\\ 
  \hline 
   2.5&  e    & 2.5 &  e  & 211640.65(8)   & 6 & 0.201 & -- & -- & --& --\\ 
   3.5&  e    & 2.5 &  e  & 211658.09(8)   & 8 & 3.010 & 10.25(0.20)  &  3.76(0.21)  & 3.67 & 14.70(0.16) \\ 
  2.5&  e    & 1.5 &  e  & 211670.03(8)  & 6 & 2.810 & $-$6.90(0.30) & 3.76(0.21) & 2.72 & 10.58(1.64)\\ 
\hline
    \end{tabular}
    \label{tab:spec_properties}
    \tablefoot{The rest frequencies ($\nu$) are taken from the laboratory measurements of \citet{Steenbakkers2025}; the corresponding experimental uncertainties are given in parentheses. The velocity scale of the spectra were aligned to the frequency of the strongest HFS line for both doublets.}
    }
\end{table*}

\section{Observations} \label{sec:observations}
The search for the lowest-lying $J= 3 \rightarrow 2$ rotational ground state transition of \cthp\ near 211~GHz,
uniquely characterized by its $\Lambda$-doublet each with  three HFS components (Table~\ref{tab:spec_properties}),
was carried out using the PI230 receiver and the low frequency module of the new First Light APEX Submillimetre Heterodyne Receiver, nFLASH230, on the Atacama Pathfinder EXperiment (APEX) 12~m sub-millimetre telescope \citep{Gusten2006}. 
The pointed observations were carried out across multiple epochs between 2021 and 2023, under project id: M9515C\_108. The half-power beam width (HPBW) is 29$^{\prime\prime}$ at 211~GHz. The spectra were converted from antenna temperature ($T_{\rm A}^*$) scales to main-beam temperature ($T_{\rm MB}$) scales using a forward efficiency of 0.95 and main beam efficiencies of ${\sim\!0.66}$ and 0.84 were assumed for the PI230 and nFLASH230 receivers, respectively. The observations were carried out in position switching mode with a relative offset of ($-$600$^{\prime\prime}$, 0$^{\prime\prime}$) in equatorial coordinates (akin to \citealt{Cuadrado2015}). 
The MPIfR-built Fourier Transform Spectrometer (FFTS4G) backends \citep{Klein2012}, which cover bandwidths of 7.8~GHz in each sideband, provided a spectral resolution of 61~kHz corresponding to a native velocity resolution of 0.086~km~s$^{-1}$. The observations were performed under moderate weather conditions when the precipitable water vapour column was between 1.5 and 3.5~mm, which for this frequency window results in atmospheric transmission levels above 80\!\%. 

The data were subsequently reduced and processed using the GILDAS/CLASS software\footnote{Software package developed by IRAM; see \url{https://www.iram.fr/IRAMFR/GILDAS/} for more information regarding the GILDAS package.} \citep{Pety2005gildas}. Polynomial baselines up to  second order were removed, and the resulting spectra were box-smoothed to a channel width of 0.1~km~s$^{-1}$. The final spectrum achieves a root-mean-square noise level of 0.77~mK at a velocity resolution of 0.1~km~s$^{-1}$ after a total integration time of 118.6~hours.\\

\begin{figure*}
    \centering
    \includegraphics[width=1\textwidth]{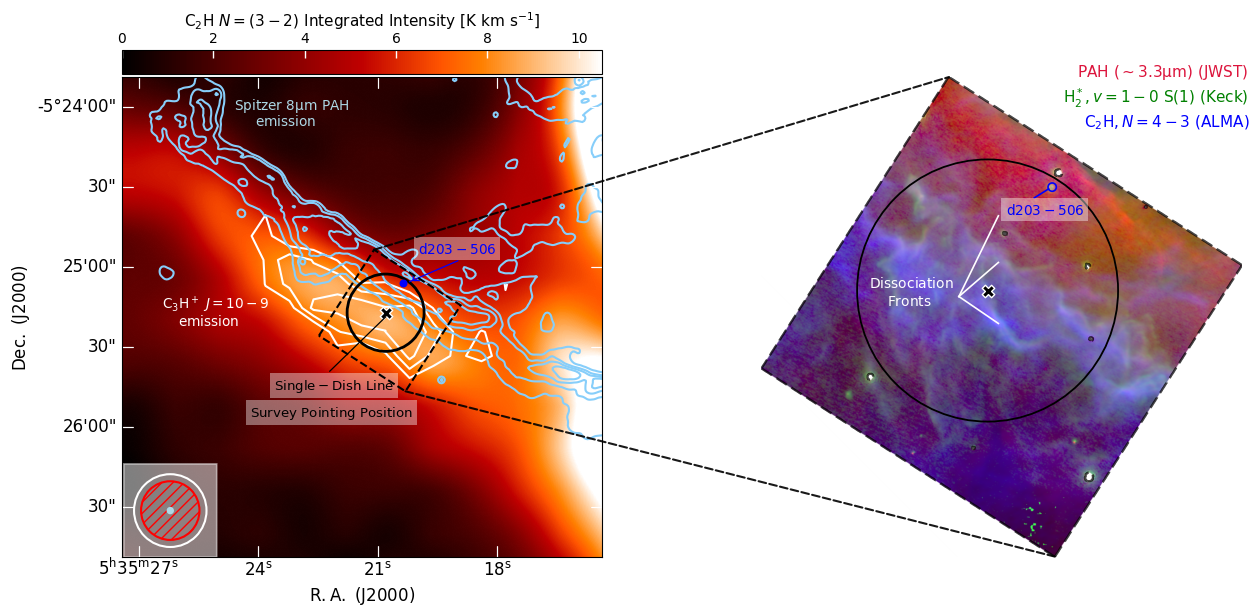}
    \caption{Left: Contours of the $l$-C$_3$H$^+$ $J=10$--9 integrated line intensity (in white) at levels of $0.97$, $1.07$, and $1.17~\mathrm{K~km~s^{-1}}$, together with Spitzer 8~$\mu$m PAH emission (in blue) shown at 3000, 4000, 4900, and 5 700~MJy~sr$^{-1}$ \citep{Walmsley2000}, are overlaid on the C$_2$H $J=3$--2 integrated intensity map (5--15~K~km~s$^{-1}$, colour scale; \citealt{Brinkmann2020}) toward the Orion Bar. Beam sizes of the different observations are indicated in the lower left corner. The position of the targeted search is marked with a black ‘$\times$’, and the beam size of the present observations is shown as a black circle. The blue circle indicates the position of the proplyd d203$-$506, toward which CH$_3^+$ has been detected \citep{Berne2023}. Right: Zoom-in over a $40^{\prime\prime} \times 40^{\prime\prime}$ field encompassing the APEX 12~m beam at the nominal single-dish line survey pointing position. This field of view is also marked by the dashed box in the left-hand panel. The inset shows a sub-arcsecond view of the Orion Bar, with $\sim 3.3~\mu$m PAH emission traced by JWST F335M–F330M in red \citep{Peeters2024}, H$_2^*$ $v=1$--0~S(1) emission observed with Keck in green \citep{Habart2023}, and C$_2$H $N=4$--3 from ALMA in blue \citep{Goicoechea2025}, highlighting the complex network of dissociation fronts embedded within the region.
    }
    \label{fig:Orion-bar-C3Hp-overlay}
\end{figure*}

The observations were conducted toward the Orion Bar, a prototypical PDR within the Orion giant molecular cloud complex. Irradiated by OB stars in the Trapezium cluster and located at a distance of ${\sim\!414}$~pc \citep{Menten2007}, its nearly edge-on orientation makes it an ideal source for studying PDR physics and chemistry. The Orion Bar has been the subject of various mm-wave and IR spectral line surveys \citep{Leurini2006, vdWiel2009, Nagy2017, Cuadrado2015, Peeters2024} and has been shown to host warm ($T_{\rm kin} = 85~$K) PDR chemistry \citep{Goicoechea2011, Berne2023}. Although the Bar is not a homogeneous structure and is comprised of corrugated structures with multiple small scale H$_2$ dissociation fronts, as revealed by high spatial resolution (1$^{\prime\prime}$) ALMA observations and more recent JWST observations \citep[][and references therein]{Goicoechea2025}, distinct PDR layers can still be identified at modest angular resolutions ($20^{\prime\prime}$--$30^{\prime\prime}$) along the interface between the molecular cloud and the H~{\small II} region (illustrated also in Fig.~\ref{fig:Orion-bar-C3Hp-overlay}). Our observations specifically targeted a position at $\alpha_{2000}$ = 05$^{\rm h}$35$^{\rm m}20\rlap{.}^{\rm s}$80, $\delta_{2000}$ =  $-05^{\circ}25^{\prime}17\rlap{.}^{\prime\prime}0$, which corresponds to the dissociation front where CO$^+$ emission peaks \citep{Stoerzer1995}. This position has been the focus of numerous molecular ion studies in the Orion Bar, leading to the first interstellar detections of species such as CF$^+$ \citep{Neufeld2006}. As a result, it has recently been dubbed the `single-dish line survey position' by \citet{Goicoechea2025}. A subset of the spectral line coverage observed toward this region as a part of this work is presented in Fig.~\ref{fig:sub-set-spec} to highlight its chemical richness.

While these discoveries support our choice of this observing position, confirming that it is the optimal pointing position for detecting \cthp\ remains difficult due to the weak nature of its emission. To address this, we mapped the spatial distribution of C$_3$H$^+$, a chemically associated molecular ion (C$_2$H$^+$ + CH $\rightarrow$ C$_3$H$^+$ + H), along the Orion Bar. Using the on-the-fly observing mode, we mapped the C$_3$H$^+$ ($J=10$\,--\,9) line emission at 224.8683~GHz with the nFLASH230 receiver. A $153^{\prime\prime} \times 36^{\prime\prime}$ map was created, centred on the CO$^+$ peak and oriented at a 35$^\circ$ angle, with an OFF position at ($-600^{\prime\prime}, 0^{\prime\prime}$) relative to the map centre. The resulting map is shown as an overlay of contours in the left-hand panel of  Fig.~\ref{fig:Orion-bar-C3Hp-overlay}.

While these observations do not spatially
resolve the emission arising from the different dissociation fronts seen at small scales, it is clear that the C$_3$H$^+$ emission is found to lie below the peak of the PAH emission, as traced by the IRAC 8~$\mu$m band (taken from the Spitzer archive), but is closely aligned with the molecular gas ridge, traced by C$_2$H ($N=3$\,--\,2) emission near 262~GHz taken from \citet{Brinkmann2020} and at the interface of the FUV pumped, vibrationally excited H$_2$ \citep{Walmsley2000}. These observations validate our choice of pointing position, assuming that \cthp\ and C$_3$H$^+$ are chemically linked, as the peak in C$_3$H$^+$ emission (at the given spatial resolution) coincides with the CO$^+$ emission peak. Additionally, the overlap with the proplyd position (at $\alpha_{2000}$ = 05$^{\rm h}$35$^{\rm m}20\rlap{.}^{\rm s}36$, $\delta_{2000}$ = $-05^{\circ}25^{\prime}05\rlap{.}^{\prime\prime}81$), where CH$_3^+$ has been recently detected in the Orion Bar \citep{Berne2023}, further supports our choice of pointing position.

\section{Results} \label{sec:results}

Figure~\ref{fig:C2Hp_spec} displays the detected HFS  lines of both $\Lambda$-doubling components of the $J= 3 \rightarrow 2$ ($\Omega=2$) rotational transition of \cthp\ towards the CO$^+$ peak in the Orion Bar. The corresponding line parameters derived from the fits are listed in Table~\ref{tab:spec_properties}. The methodology used to perform the spectral line fitting and to derive column densities is described in Sect.~\ref{subsec:spec-coldens}, followed by a discussion of photochemical modelling carried out with the Meudon PDR code in Sect.~\ref{subsec:PDR-model}, which places the inferred hydrocarbon column densities in the context of gas-phase chemistry.

\subsection{Spectral line analysis and column densities}

\label{subsec:spec-coldens}

\begin{figure}
    \centering
    \includegraphics[width=1\linewidth]{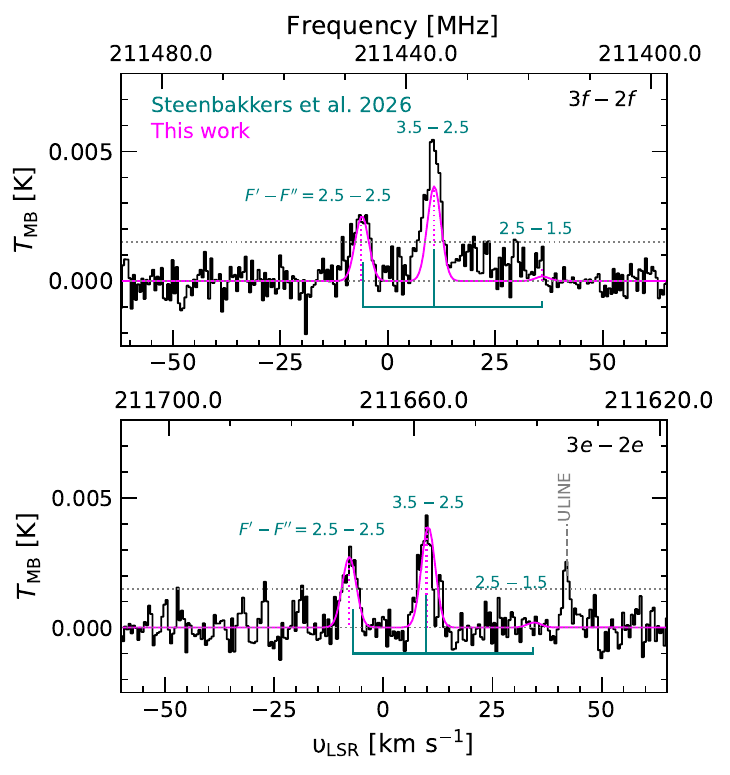}
    \caption{Spectra of the HFS-resolved $\Lambda$-doubling components (upper and lower panel)
    of the $J = 3 $--$2$ ($\Omega=2$) rotational transition of \cthp\ 
    toward the Orion Bar single dish line survey position.
    The observation  is given in black, with the GILDAS-Weeds model fit assuming $T_{\rm rot}=14~$K and a source size of 20$^{\prime \prime}$ overlaid in magenta. The relative intensities of the HFS components are marked in teal as determined by \citet{Steenbakkers2025}. The velocity scale was defined by adopting the rest frequency of the strongest HFS component of each $\Lambda$-doublet transition. With this convention, the centroid velocity of the strongest component is measured at $v_{\rm LSR}=10.25$~km~s$^{-1}$. The 3$\sigma$ detection threshold is marked by the horizontal grey dotted line.}
    \label{fig:C2Hp_spec}
\end{figure}

The spectral lines are modelled using Weeds \citep{Maret2011}, an extension of the GILDAS-CLASS software designed to analyse and fit spectral line observations under conditions of local thermodynamic equilibrium (LTE). Because only a single rotational transition of \cthp\ was detected, its rotational temperature and column density cannot be constrained independently. We therefore extracted the line profile parameters of chemically related species, that is, other small hydrocarbons also detected in our data to constrain the plausible range of rotational temperatures for \cthp. This approach assumes that the detected small hydrocarbons trace the same UV-irradiated gas within the 29$^{\prime\prime}$ APEX beam at 211~GHz. While their spatial distributions may differ at higher angular resolution, such variations are averaged over the beam and are therefore not resolved by our observations. Although C$_2$H was also observed, the spectral coverage included only its {$N=3$--2} transition, which alone is insufficient for constructing a rotational diagram.
Multi-Gaussian profiles were fitted to all detected transitions of C$_3$H$^+$, $l$-/ $c$-C$_3$H, C$_4$H, and $l$-/ $c$-C$_3$H$_2$ using Weeds. The corresponding spectral line fits and derived parameters are shown in Fig.~\ref{fig:spec} and summarised in Table~\ref{tab:other_lines}. Most spectral lines are well described by a single velocity component peaking, on average, at $\sim$10.8~km~s$^{-1}$ with line widths of $2.0 \pm 0.4$~km~s$^{-1}$. However, transitions of several species required an additional component centred near 10.2~km~s$^{-1}$ with broader line widths of (on average) $3.6 \pm 0.6$~km~s$^{-1}$, suggesting emission from a more extended gas component. Upper limits from non-detections were included to ensure adequate sampling for the construction of rotational diagrams, shown in Fig.~\ref{fig:rot-diag}.

The resulting rotational temperatures span a representative range of 14--138.5~K. 
Given the substantial uncertainty across the full temperature range, Weeds modelling of \cthp\ was performed using both temperature extremes to bracket the corresponding range of column densities. The corresponding Weeds model fit obtained for $T_{\rm rot}=14~$K is shown in Fig.~\ref{fig:C2Hp_spec}.
The HFS lines exhibit linewidths of $\sim$3.8~km~s$^{-1}$ and relative intensities close to those expected under conditions of local thermodynamic equilibrium (LTE). In line with these expectations, the two strongest HFS components of each $\Lambda$-doublet are detected, while the weakest component remains undetected down to a root-mean-square noise level of 0.77~mK at a velocity resolution of 0.1~km~s$^{-1}$. The feature seen at around 211635~MHz is an unidentified line (U-line); a check in standard catalogues (CDMS \citep{Muller2005}, JPL \citep{Pickett1998} and LSD \citep{Motiyenko2025}) did not provide a reasonable match. A comparison of the observed line intensities between the two $\Lambda$-doublet components indicates that the 211435.49~MHz transition is also contaminated, although the dominant source of this contamination remains unclear. Owing to the low signal-to-noise ratio of the detected \cthp\ hyperfine components, the data do not justify a decomposition into multiple velocity components. We therefore fitted a single velocity component with the hyperfine frequencies fixed to their laboratory values. The resulting centroid velocity of 10.3~km~s$^{-1}$ is consistent, within the uncertainties, with the secondary velocity component identified in several other hydrocarbon species, including C$_2$H. While we cannot exclude a contribution from other velocity components, the present data does not warrant a more complex kinematic model.

Assuming that the \cthp\ emission fills the APEX beam, the derived column densities span $(0.18$--$1.48)\times10^{11}$~cm$^{-2}$ for the adopted rotational-temperature range of 14--138~K. We next investigate the two main assumptions underlying this estimate: the emitting size and the excitation conditions.
The APEX 12~m beam at 211~GHz has a full width at half maximum of approximately 29$^{\prime\prime}$. To assess the impact of beam dilution, the Weeds analysis was repeated for a range of assumed source sizes, from emission filling the beam down to compact configurations of 5$^{\prime\prime}$. To constrain the plausible emitting size, we modelled the C$_2$H hyperfine structure assuming an excitation temperature of 26~K, as derived by \citet{Cuadrado2015}. We find that source sizes smaller than $\sim$20$^{\prime\prime}$ produce appreciable optical depths, leading to deviations from the HFS line ratios expected for optically thin emission under LTE. Assuming that \cthp\ originates from the same gas component as C$_2$H, we therefore adopt 20$^{\prime\prime}$ as a lower limit on the size of the emitting region. Similarly, in their work, \citet{Cuadrado2015} considered two limiting cases for the hydrocarbon emission in the Orion Bar: a semi-extended source of 9$^{\prime\prime}$ (equivalent to the IRAM telescope beam at 1~mm) and fully extended emission filling the telescope beam. A comparison of the C$_2$H fits obtained for different assumed source sizes is presented in Fig.~\ref{fig:C2H}, illustrating the impact of beam dilution on the derived parameters. We therefore adopt an emitting source size of 20$^{\prime\prime}$ throughout the remainder of this work. Correcting for the corresponding beam dilution increases the inferred \cthp\ column densities to $(0.32$--$2.15)\times10^{11}$~cm$^{-2}$.

\begin{figure}
    \centering
    \includegraphics[width=1\linewidth]{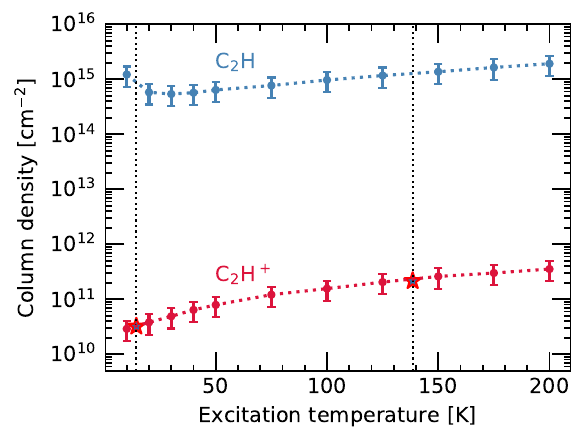}
    \caption{Variation of $N({\rm C_{2}H})$ (blue) and $N({\rm C_{2}H^+})$ (red) as a function of excitation temperature, assuming an emitting region of 20$^{\prime\prime}$. The error bars denote the formal 3$\sigma$ uncertainties derived from the non-linear least-squares fitting procedure. The "*" markers indicate the rotational temperatures adopted in our analysis ($T_{\rm rot}=14.3$ and 138.5~K) for estimating the column density of \cthp, which are also marked by the vertical dotted lines.}
    \label{fig:tex-ncol}
\end{figure}

Nevertheless, this does not fully constrain the excitation conditions or the specific gas component from which \cthp\ arises. However, it is reasonable to assume that \cthp\ traces physical conditions similar to those of C$_2$H, particularly given that our single-dish beam encompasses multiple components and dissociation fronts.
We therefore briefly examine the consequences of deriving column densities under the assumption of a fixed excitation temperature ($=T_{\rm rot}$). Modelling the column densities of C$_2$H and \cthp\ using Weeds, as described above, we find that the derived column densities increase by factors of 1.7 and 4.2, respectively, when the excitation temperature is raised from $T_{\rm rot}=20$~K to $T_{\rm rot}=100$~K.
As shown in Fig.~\ref{fig:tex-ncol}, a particularly striking result is the non-monotonic behaviour of the C$_2$H column density as a function of excitation temperature. The derived column density initially decreases with increasing values of $T_{\rm rot}$ up to $\sim$30~K, after which it gradually increases, reaching a value comparable to that at 10~K by $\sim$125~K. Beyond this temperature, it continues to increase, although more gradually. In contrast, the \cthp\ column density shows a steady increase with excitation temperature, with the growth rate flattening beyond $T_{\rm rot}\approx150$~K. Variations in the trends between \cthp and C$_2$H likely reflects the different rotational levels probed by the observations. The detected \cthp\ transitions arise from the lowest accessible rotational levels \citep[the $J=0,1$ levels are forbidden in the $\Omega=2$ ladder]{Steenbakkers2025}, whereas the $N=3$--2 line of C$_2$H is a higher-lying transition. As a result, the Boltzmann population distribution differs substantially between the two species at a given temperature.

\begin{table}[]
{\small 
    \centering
     \caption{Synopsis of the derived rotational temperatures ($T_{\rm rot}$), total column densities ($N$(X)) and abundances, assuming an emission size of 20$^{\prime\prime}$.}
    \begin{tabular}{lrll}
    \hline \hline 
        Species & \multicolumn{1}{c}{$T_{\rm rot}$} & \multicolumn{1}{c}{$N$(X)} & Abundance\tablefootmark{$\dagger$} \\
         & \multicolumn{1}{c}{[K]} &  \multicolumn{1}{c}{[cm$^{-2}$]} &   \multicolumn{1}{c}{$N$(X)/$N_{\rm H}$} \\
         \hline 
        \cthp\ & \multicolumn{1}{c}{14.3} & $(3.2 \pm 0.6)\times10^{10}$ & $5.1\times10^{-13}$\\
        & \multicolumn{1}{c}{138.5} & $(2.2 \pm 0.8)\times10^{11}$  & $3.5\times10^{-12}$ \\
        C$_2$H & \multicolumn{1}{c}{26.0\tablefootmark{*}} &  $(5.6 \pm 2.2) \times 10^{14}$ &  $8.8 \times10^{-9}$\\
        $l$-C$_3$H & $14.3 \pm 6.8$ & $(1.5 \pm 0.7) \times 10^{12}$ & $2.4\times10^{-11}$\\
        C$_4$H & $49.7 \pm 0.2$ & $(1.7 \pm 0.2)\times 10^{13}$ & $2.7\times10^{-10}$\\
        $l$-$o$-C$_3$H$_2$ & $138.5 \pm 3.5$ & $(6.5\pm 2.8)\times10^{11}\, \, $\rdelim\}{2}{*}& \multirow{2}{*}{$1.3 \times 10^{-11}$}\tablefootmark{**} \\
        $l$-$p$-C$_3$H$_2$ & $86.4 \pm 4.4$ & $(1.8\pm 0.6)\times10^{11}$\\
        $c$-$o$-C$_3$H$_2$ &  \multicolumn{1}{c}{$<91.2$} & $(4.7 \pm 1.1)\times10^{13}\, \, $\rdelim\}{2}{*}  & \multirow{2}{*}{$9.4 \times 10^{-10}$}\tablefootmark{**}  \\
        $c$-$p$-C$_3$H$_2$ &  \multicolumn{1}{c}{$<49.2$} & $(1.2 \pm 0.4)\times10^{13}$\\
         \hline 
    \end{tabular}
    \tablefoot{
   \tablefoottext{\tablefootmark{*}}{Value taken from \citet{Cuadrado2015}.}
   \tablefoottext{ \tablefootmark{$\dagger$}}{The abundance of each species with respect to H nuclei is given by $N$(X)/$N_{\rm H}$ = $N$(X)/($N$(H) + 2$N$(H$_2$)), with $N$(H)\,$\sim 3\times 10^{21}~$cm$^{-2}$ \citep{vdwerf2013} and $N$(H$_2$)\,$\sim 3\times 10^{22}$~cm$^{-2}$ derived using C$^{18}$O as a proxy \citep{Cuadrado2015}. The resulting fractional abundances should be interpreted with caution, as the \cthp\ column densities have been corrected for beam dilution assuming a source size of 20$^{\prime\prime}$, whereas the adopted H$_2$ column density is beam-averaged.} \tablefoottext{\tablefootmark{**}}{The total column densities of $l-$ and $c-$C$_3$H$_2$ are calculated as the sum of the ortho and para species, where the individual column densities were determined assuming upper limits for $T_{\rm rot}$ as noted above.}
   }
    \label{tab:col-dens}
    }
\end{table}

\subsection{Meudon PDR model}
\label{subsec:PDR-model}
As described in Sect.~\ref{sec:intro}, a long-standing puzzle in astrochemistry is the unexpectedly high observed abundances of both small and large hydrocarbons in the ISM. With the detection of \cthp, a key intermediate in the growth of chemical complexity, we investigate the origin of these elevated hydrocarbon abundances by examining the constraints that its determined abundance places on the broader chemical network.

To assess whether the inferred hydrocarbon column densities can be reproduced by gas-phase chemistry alone, with particular emphasis on the constraints provided by the estimated \cthp\ abundances, we performed photochemical modelling using version~7 of the Meudon PDR code \citep{Petit2006}. This 1D PDR code solves the FUV radiative transfer in a medium of gas and dust and computes the steady-state thermal balance and chemical structure as a function of depth into the cloud, parametrised by the visual extinction, $A_{\rm V}$ \citep{Goicoechea2007}. The attenuation of the FUV radiation field and the resulting temperature profile are then used to calculate steady-state chemical abundances for a given reaction network. While grain-surface and gas-grain exchange reactions can influence the chemistry in PDRs, they are not included in the present modelling, nor is PAH chemistry. The default chemical network does not incorporate surface chemistry, except for H$_2$ formation on grains and the adsorption and desorption processes that regulate the H$_2$ abundance. However, state-to-state reactions of vibrationally excited H$_2$ (or H$_2^*$) with C$^+$ or OH are explicitly treated \citep{Agundez2010}. 

Isobaric models were computed at a constant thermal pressure of $P_{\rm th}\!\sim\!10^{8}$~cm$^{-3}$~K \citep{Joblin2018, vdPutte2024}, allowing for density gradients. Although the thermal pressure is unlikely to be uniform across the single-dish survey position and the dissociation fronts encompassed within the beam (as is evidenced in the right-hand panel of Fig.~\ref{fig:Orion-bar-C3Hp-overlay}), with only a single pointing, we lack sufficient spatially resolved information to constrain density and temperature variations, thereby justifying the use of a single-layer model with constant thermal pressure. We further assume an incident FUV radiation field of $G_0\simeq2\times10^{4}$ in units of the Habing field \citep{Habing1968}, inferred from FUV-pumped IR-fluorescent lines by \citet{Peeters2024} and dust grain properties with an extinction-to-color index ratio, $R_{\rm V}= A_{\rm V}/E_{\rm B-V}$=5.5, consistent with the flattened extinction curve observed toward Orion by \citet{Cardelli1989}. The initial elemental abundances used were adopted from \citet{Sofia2004} and \citet{Goicoechea2021}, and a cosmic-ray ionisation rate of $\zeta_{\rm H}=10^{-16}$~s$^{-1}$ was adopted, motivated by recent revisions in the computed cosmic-ray ionisation rate by \citet{Neufeld2024} and \citet{Indriolo2026}. We note, however, that the total ionisation rate in the Orion Bar may be higher when accounting for X-ray ionisation from stellar sources in the Trapezium cluster (e.g., \citealt{Gupta2010}), as also discussed by \citet{Cuadrado2015}. 

\begin{figure}
    \centering
    \includegraphics[width=1\linewidth]{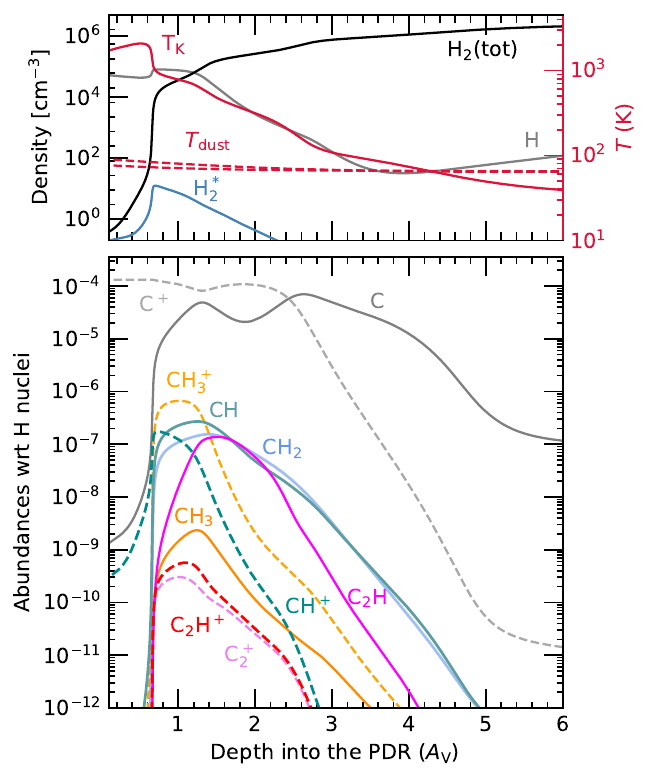}
    \caption{Isobaric PDR model with thermal pressure $P_{\rm th} = 10^{8}\,\mathrm{K\,cm^{-3}}$, computed for fixed values of $G_0 = 2 \times 10^{4}$ and $\zeta_{\rm H} = 10^{-16}\,\mathrm{s^{-1}}$. \textit{Top panel:} Gas density, gas temperature, dust temperature (shown for the minimum and maximum grain-size distributions), and the abundance of FUV-pumped H$_2^\ast$ ($v \geq 1$) as a function of depth into the cloud, expressed in terms of the visual extinction, $A_{\rm V}$. \textit{Bottom panel:} Abundance profiles of selected hydrocarbon species as a function of $A_{\rm V}$. Neutral species are shown with solid curves, while ionic species are represented by dashed lines; different colours correspond to different species as labelled.  }
    \label{fig:PDR-all}
\end{figure}

The upper panel of Fig.~\ref{fig:PDR-all} shows the physical and chemical structure of the Meudon PDR models for the Orion Bar, presenting the gas density, gas temperature, dust temperature (for both the minimum and maximum grain-size distributions), and the abundance of FUV-pumped H$_2^\ast$ ($v\!\geq\!1$) as a function of depth into the cloud, expressed in terms of visual extinction, $A_{\rm V}$. These results are consistent with the most recent Meudon PDR models of the Orion Bar presented by \citet{Goicoechea2025}.

\begin{figure}
    \centering
    \includegraphics[width=1\linewidth]{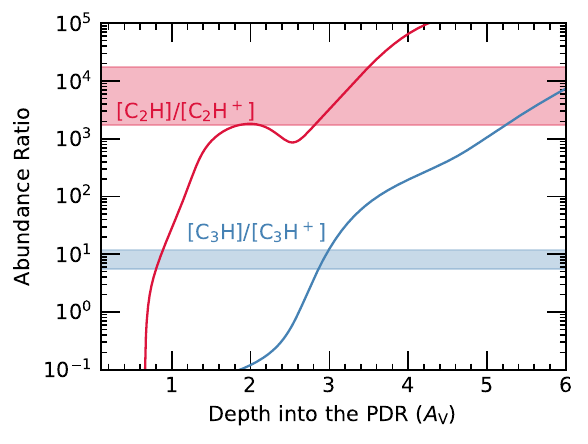}

    \caption{Meudon PDR model-predicted abundance ratios for [C$_2$H]/[\cthp] (red) and [C$_3$H]/[C$_3$H$^+$] (blue), with the corresponding ranges of observed abundance ratios in the Orion Bar shown in the same colours.}
    \label{fig:abundance-ratio}
\end{figure}

The lower panel displays the abundance profiles of gas-phase \cthp, C$_2$H, C$^+$, and C, along with other related small hydrocarbons and radicals. The rise in carbon abundance around $A_{\rm V}\!\sim\!0.7$~mag coincides with the peak of FUV-pumped H$_2^\ast$, beyond which CH$^+$ reaches its maximum, followed by CH$_3^+$ and C$_2$H$^+$. The abundance of \cthp\ peaks at $A_{\rm V} = 1.3$~mag. However, when comparing the abundance ratios [C$_2$H]/[\cthp] and [C$_3$H]/[C$_3$H$^+$] (the latter from \citealt{Cuadrado2015}), as indicated in Fig.~\ref{fig:abundance-ratio}, the observed range of ratios (highlighted by the horizontal bands) are reproduced at $A_{\rm V}$ values between 2.7 and 3.2~mag for both molecular pairs. The [C$_2$H]/[\cthp] ratio also exhibits a secondary region of agreement at $A_{\rm V}\!\sim\!1$--2~mag, which may indicate contributions from more diffuse gas components. However, given the uncertainties associated with the adopted rotational temperatures, the derived column densities, and the limited spatial resolution of our single-dish observations, we do not attempt a more detailed interpretation of this feature. Overall, the inferred extinction range should therefore be regarded as representative of the beam-averaged gas sampled by our observations rather than of a single, well-defined PDR layer. The following section discusses the gas-phase formation pathways of \cthp\ and other small hydrocarbons in this context.

\section{Discussion} \label{sec:discussion}

\subsection{Gas-phase formation of \texorpdfstring{\cthp}{C2H+} and other hydrocarbons}
In strongly UV-irradiated PDRs, such as the Orion Bar, gas-phase carbon chemistry is initiated by the reaction C$^+$ + H$_2 \rightarrow$ CH$^+$ + H, which dominates over the slower radiative association channel, C$^+$ + H$_2 \rightarrow$ CH$_2^+$ + photon. The former reaction is highly endothermic ($E/k \approx 4620$~K; \citealt{Hierl1997}) and therefore requires either hot gas (a few hundred Kelvin or more) or FUV-pumped vibrationally excited H$_2$ to proceed efficiently \citep[e.g.,][]{Black1987, Agundez2010}. Such conditions prevail at the edge of strongly irradiated regions like the Orion Bar, studied here. When H$_2$ is vibrationally excited ($v\!\geq\!1$), the reaction barrier is effectively overcome, enabling rapid formation of CH$^+$ \citep{Godard2013, Faure2017}.
The production of CH$^+$ constitutes a bottleneck in the hydrocarbon reaction network \citep{Zannese2025, Goicoechea2025}, however, once formed, CH$^+$ undergoes barrier-less hydrogen abstraction reactions with H$_2$ to produce CH$_2^+$ and CH$_3^+$. These simple hydrocarbon ions subsequently recombine, leading to the formation of CH and other small hydrocarbons. In regions with gas temperatures above 500~K and strong FUV fields --such as in our slab model, where $T_{\rm k} > 500$~K across the relevant $A_{\rm V}$ ranges for \cthp\ excitation-- H$_2$ excitation is sufficient to sustain efficient CH$^+$ formation and, consequently, enhanced hydrocarbon abundances. Under these conditions, CH$^+$ may also form in excited states or undergo chemical pumping, a mechanism invoked to explain the observed line intensities of C$_2$H \citep{Faure2017, Goicoechea2025}. In contrast, in low-FUV PDRs such as the Horsehead \citep{Pety2005}, where these energetic conditions are not met, CH$^+$ and related species are far less prominent. In this framework, CH$^+$ formation marks the onset of hydrocarbon chemistry, followed by successive reactions that ultimately lead to the formation of more complex species (see Fig.~\ref{fig:chemical_network}).

\begin{figure*}
    \centering
    \includegraphics[width=0.9\linewidth]{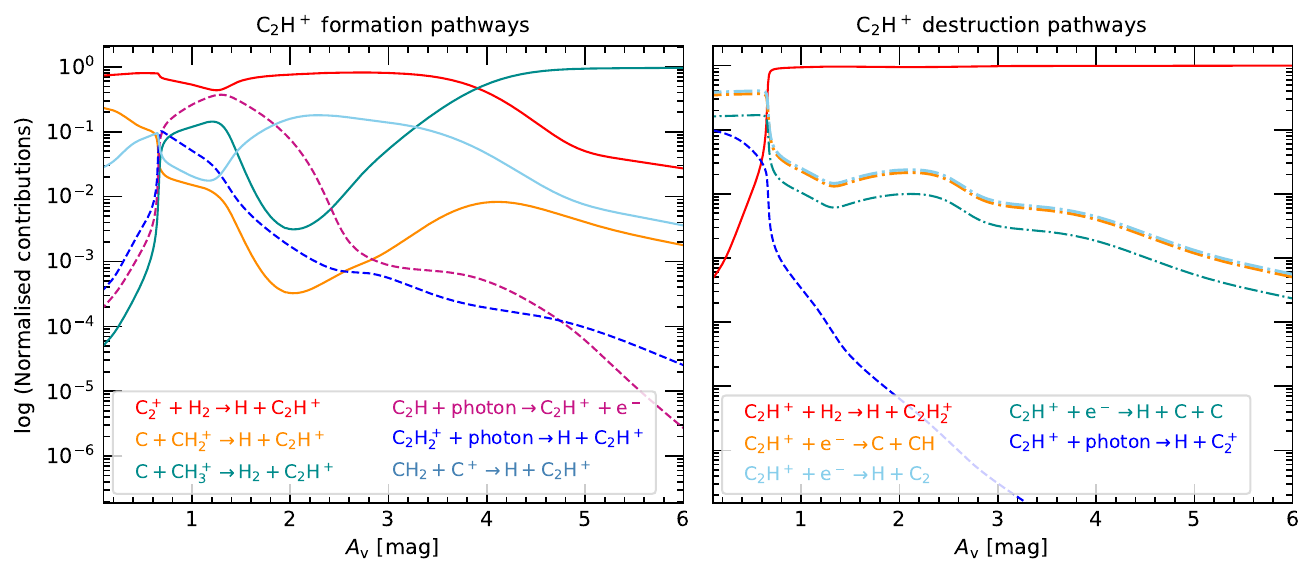}
    \caption{Normalised contributions of the  formation (left) and destruction (right) pathways for \cthp\ for the physical conditions in the Orion Bar modelled with the Meudon PDR code. In both panels the dominant reaction pathway for $A_{\rm V}$ ranges of interest (>1~mag) are displayed in red. All ion-neutral reactions are represented by solid curves, while photodissociation and dissociative recombination reactions with electrons are marked by dashed and dashed-dotted curves, respectively.  }
    \label{fig:formation-destruction}
\end{figure*}

To understand the role of \cthp\ in the broader astrochemical context, we here analyse the primary formation and destruction pathways of \cthp\, for the physical conditions modelled above (Fig.~\ref{fig:formation-destruction}). It is found that hydrogen abstraction reactions involving C$_2^+$ appear to dominate across the full range of modelled $A_{\rm V}$. The chemistry of C$_2^+$ is generally expected to be associated with UV-irradiated, partially molecular gas in which C$^+$ remains abundant \citep{Federman1989}. In such environments, C$_2^+$ can form efficiently through reactions such as
C$^+$ + CH $\rightarrow$ C$_2^+$ + H, 
after which it rapidly undergoes hydrogen abstraction reactions with H$_2$ to form C$_2$H$^+$ (see Fig.~\ref{fig:chemical_network}).
Because C$_2^+$ reacts efficiently with H$_2$, its abundance is expected to peak in the transition layers between predominantly atomic and fully molecular gas, characteristic of strongly FUV-irradiated PDRs.

In addition, given that C$^+$ is the dominant carbon-bearing species at the PDR edge, we would also expect \cthp\ chemistry to be partially driven by reactions involving CH$_2$ and C$^+$ (indicated in light blue in Fig.~\ref{fig:formation-destruction}, where CH$_2$ has been observed toward the Orion Bar with an abundance of the order of ${\sim\!10^{-8}}$ with respect to H nuclei) and by the photoionization of C$_2$H. The former, CH$_2$ stems from the initial dissociative recombination of CH$_3^+$. The detection of H$_2^*$ \citep{Zannese2025, Goicoechea2025}, CH$_3^+$ \citep{Zannese2025}, and CH$_2$ \citep{Jacob2021} near the dissociation front, together with C$_2$H \citep{Nagy2017, Cuadrado2015, Goicoechea2025} and \cthp\ in the same regions (albeit not necessarily traced on directly comparable spatial scales) suggests that all the necessary ingredients are in place, lending growing observational support to a bottom-up picture of hydrocarbon chemistry, with the chemical ``ingredients'' seemingly assembled and ready for molecular complexity to unfold.\\

At cloud depths $1 \lesssim \! A_{\rm V}\! \lesssim\!3$~mag, where the abundance of C$_2$H$^+$ peaks in our models, its destruction is dominated by reactions with H$_2$, while dissociative recombination dominates closer to the PDR surface. This hydrogen abstraction reaction is assumed to be barrierless and produces C$_2$H$_2^+$ \citep{herbst1993calculations}, which can further react with H$_2$ to form C$_2$H$_3^+$ (this reaction is endothermic, and thus enhanced in warm gas environments with H$_2^*$), which upon electron recombination then yields neutral acetylene (C$_2$H$_2$). The resulting sequence links the chemistry of \cthp\ directly to the production of one of the key building blocks of interstellar hydrocarbon growth.

Beyond serving as a precursor of neutral acetylene, C$_2$H$_2^+$ and related small hydrocarbon ions are themselves highly reactive and can efficiently undergo ion–molecule reactions with unsaturated hydrocarbons, providing pathways toward larger molecular species under interstellar conditions \citep{schmid2020isomer}. Likewise, neutral acetylene readily reacts with hydrocarbon ions, leading to carbon-chain growth and potentially contributing to the formation of aromatic species and PAHs \citep{rap2022low}. These ion-driven routes may be particularly effective in strongly UV-irradiated gas where ionic abundances remain high.
Neutral acetylene can also participate in the hydrogen abstraction–acetylene addition (HACA) mechanism, in which a hydrogen atom is first abstracted from a hydrocarbon, creating a reactive radical site that subsequently undergoes sequential addition of C$_2$H$_2$ enabling the stepwise formation of larger hydrocarbons and PAHs \citep{Frenklach1989}. Barrier estimates for acetylene addition ($E$ = 1200--2400~K) suggest that low temperatures would normally inhibit HACA. However, in regions like the Orion Bar, the same physical conditions that allow CH$^+$ formation --warm gas and vibrationally excited H$_2$-- may also render HACA energetically feasible. Complementary mechanisms such as hydrogen abstraction–vinylacetylene addition (HAVA) may further contribute to the growth of larger benzoid-PAHs \citep{Zhao2018}.

\begin{figure*}
    \centering
    \includegraphics[width=0.9\linewidth]{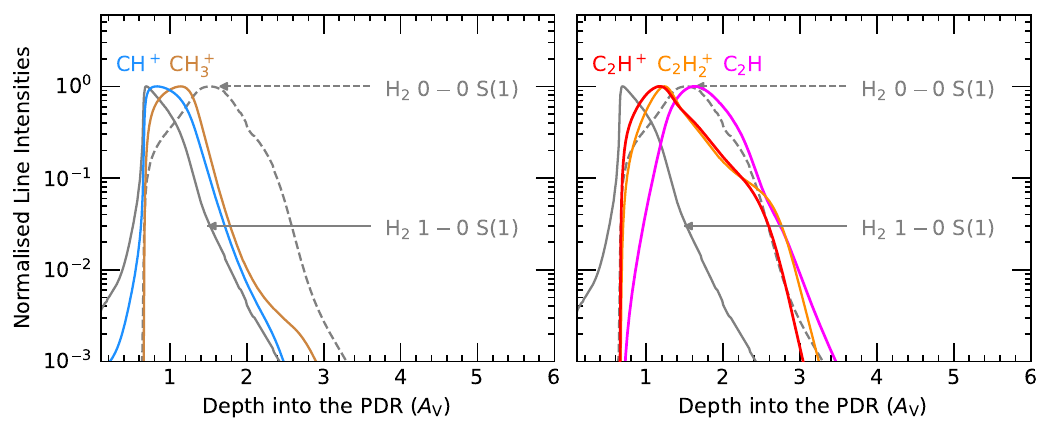}
    \caption{Normalised line intensities of H$_2$ (0--0) S(1) (dashed grey) and H$_2$ (1--0) S(1) (solid grey) alongside that of CH$^+$ (blue) and CH$_3^+$ (brown) on the left-hand panel and that of \cthp\ (red), C$_2$H (magenta) and C$_2$H$_2^+$ (dark orange) for the physical conditions in the Orion Bar modelled with the Meudon PDR code. }
    \label{fig:norm-abundances}
\end{figure*}

Figure~\ref{fig:norm-abundances} displays the normalised abundances of \cthp, CH$^+$, CH$_3^+$, and excited H$_2$. Two H$_2$ transitions are considered: the low-rotational line H$_2$ (0-0) S(1) ($E= 1015~$K), and the highly excited, FUV-pumped ro-vibrational line H$_2$ (1-0) S(1) ($E = 6952~$K). The peaks of CH$^+$ and CH$_3^+$ closely align with the highly excited H$_2$, a result confirmed by high spatial resolution maps of the Orion Bar \citep[e.g.,][]{Zannese2025}, consistent with the requirement of vibrationally excited H$_2$ for their formation. In contrast, C$_2$H$_2^+$ and \cthp\ peak deeper in the cloud, near the lower-excitation H$_2$, indicating that their emission arises just below the FUV-excited layers and immediately above the denser molecular gas traced by C$_2$H. Although current observations cannot resolve this boundary and high-resolution maps are not yet feasible, it can nonetheless be conceptually visualised as a gas layer framing the C$_2$H emission presented in Fig.~\ref{fig:Orion-bar-C3Hp-overlay}. Further constraints cannot be placed on our models, as neutral acetylene has been detected in space \citep[e.g.,][]{Kanwar2024}, but not in the Orion Bar, while its cation, C$_2$H$_2^+$, remains observationally elusive, although laboratory studies have characterised its low-energy vibrational modes \citep{Asvany2005,Steenbakkers2024} and its CH-stretching transition in the 3~$\mu$m region \citep{Jagod1992,Schlemmer2024}. In this context, the detection and modelling of C$_2$H$^+$ in the Orion Bar provides a key constraint on early ion–molecule hydrocarbon chemistry. It traces the initial reaction steps leading to C$_2$H$_2^+$ and C$_2$H$_3^+$, which are important intermediates in the formation of neutral acetylene among other interstellar species.

\subsection{\texorpdfstring{\cthp}{C2H+} in other regions}
The detection of PAHs in space has long been associated with warm, carbon-rich environments, where their formation is typically linked to circumstellar envelopes around evolved stars \citep{Leger1984, Allamandola1985, Cherchneff1992}. This picture became considerably more intriguing with the discovery of aromatic chemistry in the cold, UV-shielded pre-stellar core TMC-1 \citep{McGuire2018, Cernicharo2021}. A growing inventory of PAHs and related aromatic species has been uncovered in TMC-1, enabled largely by the extensive GOTHAM \citep{McGuire2018} and QUIJOTE \citep{Cernicharo2021a} line surveys, which include for example, 1- and 2-cyanonaphthalene \citep{McGuire2021}, indene \citep{Cernicharo2021b, Burkhardt2021}, and most recently cyanocoronene \citep{Wenzel2025} and phenalene \citep{Cabezas2025}. Collectively, these detections demonstrate that aromatic ring formation can proceed in gas at $\sim$10~K, well shielded from external UV radiation. Because relatively small PAHs are unlikely to survive the earlier diffuse-cloud phase, these findings strongly argue for \textit{in situ} bottom-up formation within TMC-1 itself \citep{Agundez2021}.

A defining feature of this cold chemistry is the presence of abundant small unsaturated hydrocarbons and radicals. The propargyl radical (CH$_2$CCH), detected at high abundance relative to H$_2$, is a well-established intermediate in benzene formation \citep{Agundez2021}. Vinyl acetylene (CH$_2$CHCCH), allenyl acetylene (H$_2$CCCHCCH), and propylene (CH$_2$CHCH$_3$) are also present in significant quantities \citep{Cernicharo2021b, Marcelino2007}. Together, these species provide plausible pathways to first-ring closure through radical–neutral and ion–neutral reactions \citep{Mallo2025}, even at very low temperatures. Laboratory and theoretical studies indicate that several of these reactions proceed either without an energy barrier or via tunnelling assistance \citep{Woon2002}, rendering bottom-up growth chemically viable in cold dark clouds.

If aromatic chemistry indeed proceeds bottom-up in TMC-1, then constraining the column densities and chemical role of small ions such as \cthp\ becomes essential \citep{McGuire2018, Cernicharo2021c}. Their detection would provide direct evidence that ion-driven growth pathways operate efficiently even at 10~K, reinforcing the idea that complex hydrocarbons do not require strong UV fields or high temperatures to emerge.

\begin{figure}
    \centering
    \includegraphics[width=0.95\linewidth]{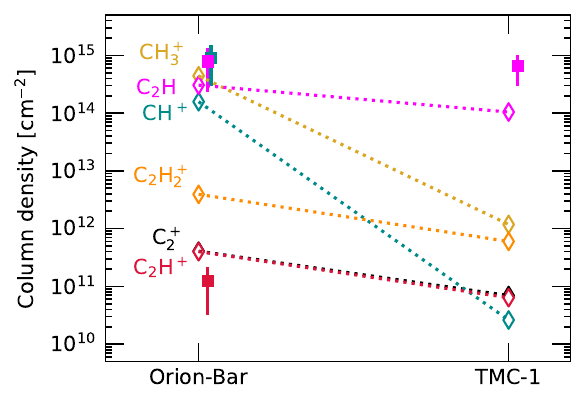}
    \caption{Model-predicted column densities from the Meudon PDR code (open diamonds) for \cthp\ and related small hydrocarbons, including C$_2^+$, toward the Orion Bar and TMC-1 compared with column densities determined from observations (filled squares). Error bars indicate the range between extended and compact source-size assumptions. The models assume a face-on PDR geometry with $A_{\rm V,tot} = 10$~mag.}
    \label{fig:Orion-TMC}
\end{figure}

To assess whether \cthp\ can be sustained under such conditions, we computed Meudon PDR models adopting the physical parameters reported for TMC-1 in face-on geometry ($G_0 = 6.5$, $\zeta_{\rm H} = 10^{-16}$~s$^{-1}$, and $P_{\rm th} = 1.5\times10^{4}$~cm$^{-3}$~K) \citep{Millar1984, Ebisawa2019, Fuente2019}. Figure~\ref{fig:Orion-TMC} presents the predicted integrated column densities (i.e., total column density of each species through the modelled cloud) from the Meudon code of simple hydrocarbon species compared with observed values where available. Overall, the model reproduced column densities are higher in the Orion Bar than in TMC-1. For C$_2$H, the model reproduced column densities in the Orion Bar lie within observational uncertainties (with the notable exception of CH$^+$), whereas in TMC-1 they are significantly underestimated \citep{Cernicharo2022}. This discrepancy may suggest that the adopted value for the cosmic-ray ionization rate is too low, thereby limiting the abundance of H$_{3}^{+}$ and related molecular ions (e.g., HCO$^+$) that sustain ion-molecule reaction pathways in TMC-1 \citep{Majumdar2017, Fuente2019}, potentially suppressing the formation of hydrocarbon ions and larger hydrocarbons in cold cores.

The most striking differences relative to the Orion Bar are found for CH$_3^+$ and CH$^+$. This is expected because, in TMC-1 there is no significant reservoir of vibrationally excited H$_2$, and thus the endothermic reaction C$^+$ + H$_2$ $\rightarrow$ CH$^+$ + H is inefficient. Instead, the ion chemistry is sustained primarily by cosmic-ray ionization, with CH$^+$ forming indirectly through reactions involving C with H$_3^+$ or channels linked to HCO$^+$ (as reproduced by the Meudon PDR models). In the TMC-1 models, the \cthp\ column density peaks at relatively low extinction ($A_{\rm V} \sim 0.5$~mag). At this depth, its formation is dominated by reactions involving C$_2^+$, with less than 25\% contribution from alternative channels such as CH$_2^+$ + C at other visual extinctions. The predicted C$_2^+$ column densities are comparable to those of \cthp\ in both the Orion Bar and TMC-1, with slightly higher column densities toward TMC-1.

In contrast, the Orion Bar represents a radically different chemical regime. There, intense FUV radiation and elevated gas temperatures drive chemistry through vibrationally excited H$_2$ and efficient CH$^+$ production. Subsequent barrierless ion–molecule reactions rapidly build small hydrocarbons such as C$_2$H and C$_3$H$^+$. Under these energetic conditions, reaction barriers that are prohibitive in cold clouds are readily overcome.

Despite these environmental differences, an important commonality emerges: in both regions, molecular growth appears to be initiated by small, highly reactive intermediates. In TMC-1, where neutral–neutral channels are largely suppressed by low temperatures, ion–molecule chemistry becomes particularly significant. Small hydrocarbon cations such as \cthp\ may therefore play a central role in coupling abundant C$_2$ building blocks (e.g., C$_2$H$_2$) to larger unsaturated chains and, ultimately, cyclic structures. As discussed in the previous section, through reactions with H$_2$, \cthp\ can form C$_2$H$_2^+$ and C$_2$H$_3^+$, which upon dissociative recombination regenerate acetylene, a fundamental precursor for aromatic chemistry. 
With predicted column densities on the order of $\sim\!6.4 \times 10^{10}$~cm$^{-2}$, within the range of uncertainties associated with our modelling of \cthp\ in the Orion Bar, and contingent on the adopted physical conditions for TMC-1, its presence in cold gas cannot be excluded. As revealed by our models, even at $10$~K, cosmic-ray ionization sustains a low but persistent ion population, enabling the formation of small hydrocarbon cations. Furthermore, as discussed above the underestimation of C$_2$H in our models may indicate that the adopted cosmic-ray ionization rate in TMC-1 is too low, thereby suppressing ion–molecule reaction pathways. If so, the abundances of \cthp\ and other hydrocarbon ions may also be underestimated. Given the sensitivity reached in recent deep spectral surveys of TMC-1, a targeted search for \cthp\ may therefore be feasible. Even a stringent upper limit would provide valuable constraints on the efficiency of ion–molecule chemistry and its role in cold bottom-up hydrocarbon growth.

\section{Conclusions} \label{sec:conclusions}

We report the first interstellar detection of \cthp\ toward the Orion Bar, at the single-dish line survey position. The detection is based on APEX 12~m sub-mm observations of the lowest-lying $J=3$--$2$ ($\Omega=2$) rotational transition near 211~GHz, which provides a unique spectral signature through its resolved $\Lambda$-doubling and hyperfine structure, as recently characterised in laboratory measurements \citep{Steenbakkers2025}. We derive \cthp\ column densities between $0.3\times10^{11}$ and $2.2\times10^{11}$~cm$^{-2}$ for excitation temperatures of 14--138~K. PDR models computed with the Meudon code reproduce the observed \cthp\ abundance within uncertainties and link its formation to that of CH$^+$ and CH$_3^+$ in the Orion Bar. In these strongly UV-irradiated regions, vibrationally excited H$_2$ overcomes the endothermic barrier to CH$^+$ formation, thereby initiating a network of rapid ion–molecule reactions. The detection of \cthp\ in the Orion Bar thus highlights the efficiency of bottom-up hydrocarbon chemistry at the PDR surface. Within this network, C$_2$H$^+$ acts as a key intermediate: it traces the propagation of carbon-chain growth and serves as a precursor to C$_2$H$_2^+$ and C$_2$H$_3^+$, which are essential species in subsequent ion–molecule reactions that ultimately feed the formation of larger hydrocarbons. In this sense, \cthp\ occupies a pivotal position linking small C$_2$ building blocks to increasing molecular complexity.
This ion-driven growth bears resemblance to the HACA mechanisms invoked in combustion chemistry, where hydrogen abstraction from a hydrocarbon is followed by sequential acetylene addition, enabling stepwise formation of larger hydrocarbons and PAHs. In the ISM, analogous ion–molecule pathways involving C$_2$H$_2^+$ and C$_2$H$_3^+$ may provide a viable bottom-up route toward larger hydrocarbon cations, with \cthp\ acting as an important intermediary despite its observational elusiveness until now.
While the Orion Bar illustrates bottom-up chemistry under highly energetic conditions, cold dark clouds such as TMC-1 may provide a complementary and perhaps more surprising laboratory. There, chemistry is sustained by cosmic-ray–driven ionization, which maintains a low but persistent ion population even at $\sim 10$~K, potentially enabling the earliest stages of aromatic growth in the absence of strong UV radiation. Detecting and characterizing \cthp\ in such environments would therefore represent a decisive test of whether the seeds of PAHs can form in the darkest and coldest regions of the ISM.
While our models do not explicitly include top-down processes such as PAH fragmentation or grain-surface chemistry, the agreement between the observations and the purely gas-phase PDR predictions strongly supports a bottom-up formation scenario in which small hydrocarbon ions drive molecular complexity in FUV-irradiated regions like the Orion Bar. More concrete analysis awaits observational constraints from other related hydrocarbon ions like C$_2$H$_2^+$ and C$_2$H$_3^+$. 

\begin{acknowledgements}
We thank the anonymous referee for their constructive feedback. The data was collected under the Atacama Pathfinder EXperiment (APEX) Project, led by the Max Planck Institute for Radio Astronomy at the ESO La Silla Paranal Observatory under project id  M9515C-108. We would like to express our gratitude to the APEX staff and science team for their continued assistance in carrying out the observations presented in this work. We are thankful to the developers of the C++ and Python libraries and for making them available as open-source software. In particular, this research has made use of the NumPy \citep{numpy}, SciPy \citep{scipy} and matplotlib \citep{matplotlib} packages. A.M.J. thanks the support of the Max Planck Gesellschaft. S.K. acknowledges support from the BMWI via DLR, project number 50OR2311, and funding from the Deutsche Forschungsgemeinschaft (DFG, German Research Foundation), project number 558818801. K.S., G.C.G. and S.B. acknowledge funding through the project “HFML-FELIX: a Dutch Center of Excellence for Science under Extreme Conditions” (with Project No. 184.035.011) of the research program “Nationale Roadmap Grootschalige Wetenschappelijke Infrastructuur,” which is partly financed by the Netherlands Organisation for Scientific Research (NWO). In addition, this work has been supported by an ERC Advanced Grant (MissIons: 101020583), by the Deutsche Forschungsgemeinschaft (DFG) via the Collaborative Research Centre 1601 (project ID: 500700252, sub-projects A1, A2, B8 and C4) and by the Deutsche Forschungsgemeinschaft (DFG, German Research Foundation) under Germany’s Excellence Strategy EXC 3037 – 533607693 - Our Dynamic Universe.
W.G.D.P.S. thanks the Alexander von Humboldt Foundation for funding through a Postdoctoral Fellowship during the time this work has been carried out. 
\end{acknowledgements}

\bibliographystyle{aa} 
\bibliography{aa61924-26}

\begin{appendix}
\onecolumn
\section{APEX spectral band}\label{appendix:APEX_full-spec} 
\begin{figure*}[h]
    \centering
    \includegraphics[width=1\linewidth]{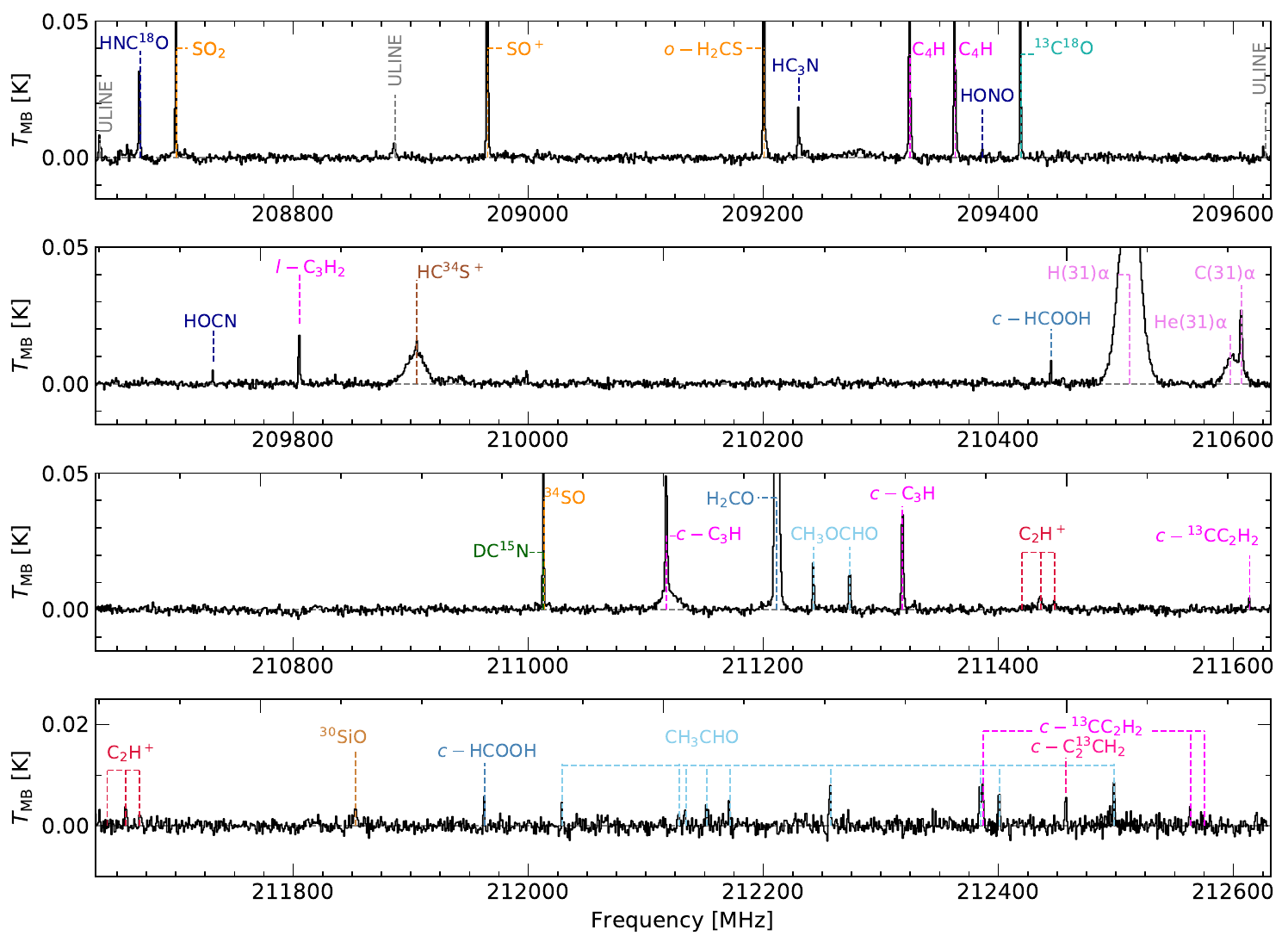}
    \caption{A subset of the baseline-subtracted spectrum obtained toward the Orion Bar single-dish line survey position, between 208632~MHz to 212632~MHz with the APEX 12~m sub-millimetre telescope in units of the main-beam temperature scale. All identified lines are labelled. }
    \label{fig:sub-set-spec}
\end{figure*}

In this Appendix, we illustrate the spectral line richness toward the Orion Bar single-dish pointing position through a spectrum, see Fig.~\ref{fig:sub-set-spec}, displaying a subset of the observed frequency band, between 208632~MHz and 212632~MHz, which also covers the \cthp\ transitions.

\newpage 
\section{APEX spectra and LTE model fit parameters of hydrocarbons}  \label{appendix:hydrocarbons}

\begin{figure*}
    \centering
    \includegraphics[width=1\linewidth]{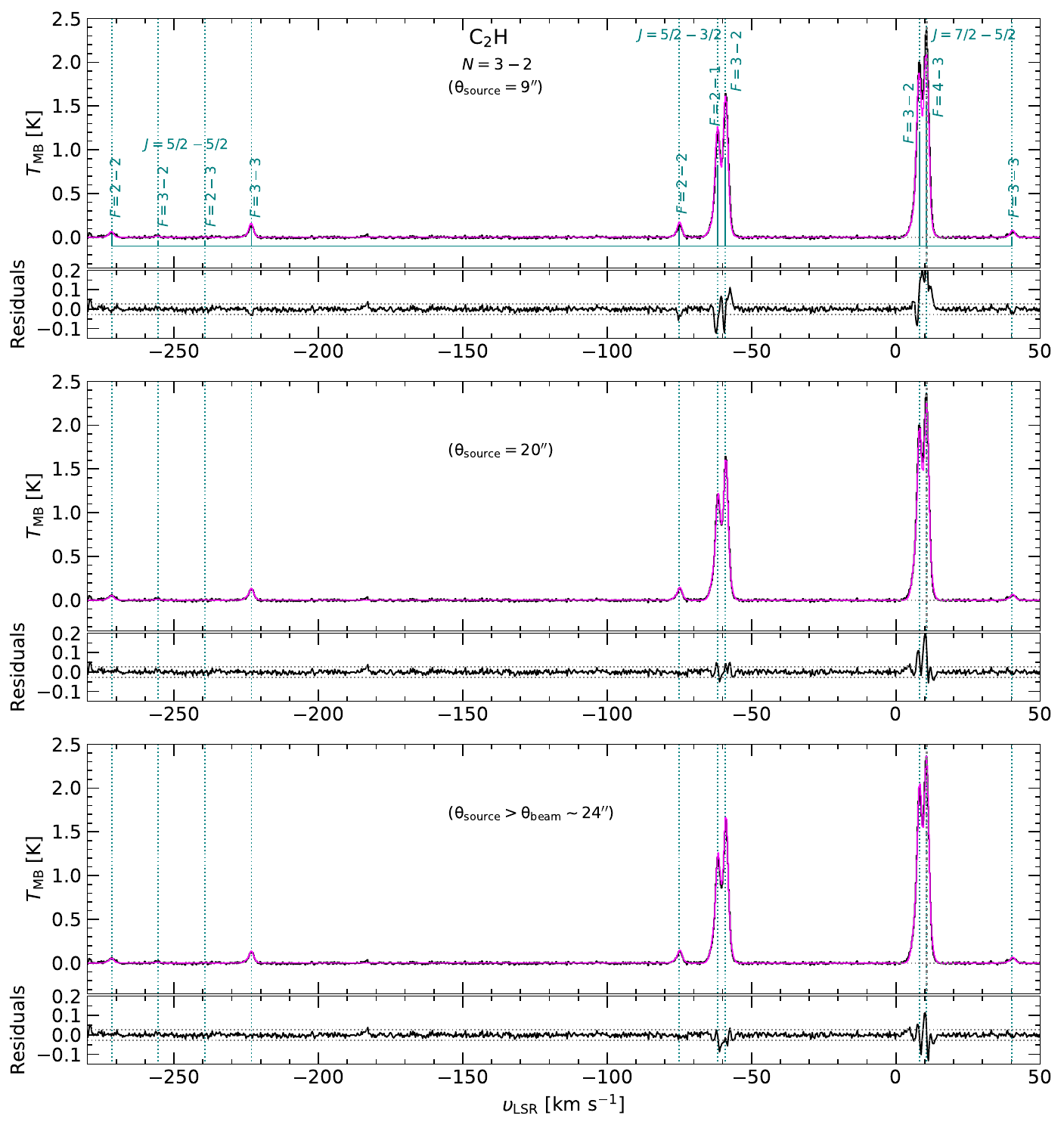}
    \caption{From top-to-bottom: Baseline-subtracted spectrum of the HFS components of the C$_2$H $N = 3$--2 transition (shown in black) toward the same position in the Orion Bar as the \cthp\ detection, overlaid with the Weeds fit (shown in magenta) assuming $T_{\rm rot}=26$~K, for a source size ($\theta_{\rm source}$) of 9$^{\prime\prime}$, 20$^{\prime\prime}$ and $>$24$^{\prime\prime}$. The positions and relative intensities of the individual HFS components are indicated in teal. The residuals of each fit highlight the impact of optical depth effects at smaller assumed source sizes.}
    \label{fig:C2H}
\end{figure*}

Table~\ref{tab:other_lines} lists the molecular line fit parameters for all detected hydrocarbons except \cthp. As discussed in the main text, jointly modelling multiple chemically related species not only strengthens the identification of \cthp\ but also improves the determination of the velocity of the gas layer from which its emission originates. Our spectral coverage includes the $J=10\rightarrow9$ transition of $l$-C$_3$H$^+$, several transitions of linear and cyclic isomers of C$_3$H and C$_3$H$_2$, as well as transitions of C$_4$H. The spectroscopic parameters for these molecules were taken from \citet{Baddeliyanage2025} for $l$-C$_3$H$^+$, \citet{Caris2009} for C$_3$H, \citet{Gottlieb1983} for C$_4$H, \citet{V1990ApJ...364L..53V} for $l$-C$_3$H$_2$, and \citet{Spezzano2012} for $c$-C$_3$H$_2$. In addition, this Appendix also presents the $N=3$--2 spectrum of C$_2$H from archival APEX data (see Fig.~\ref{fig:C2H}).

\begin{figure*}
    \centering
    \includegraphics[width=0.3\linewidth]{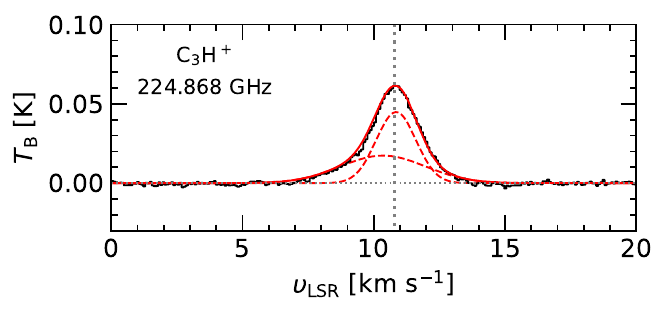}\quad
    \includegraphics[width=0.3\linewidth]{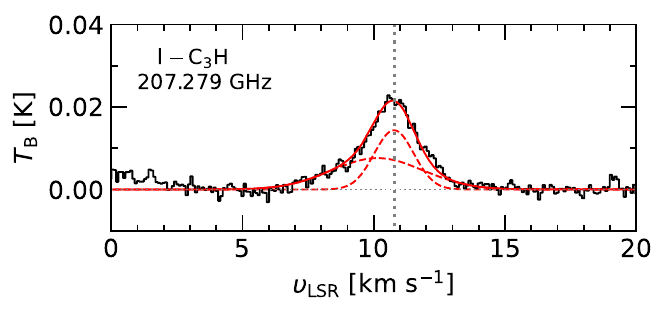}\quad
    \includegraphics[width=0.3\linewidth]{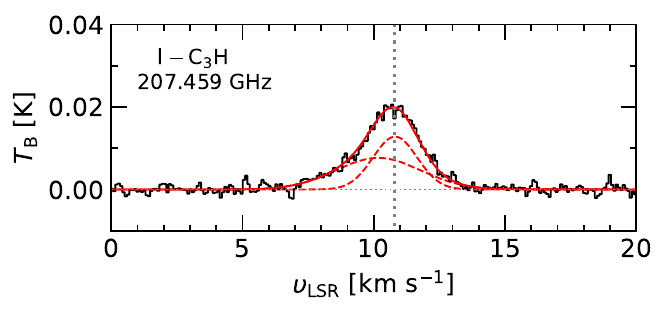}\\
    \includegraphics[width=0.315\linewidth]{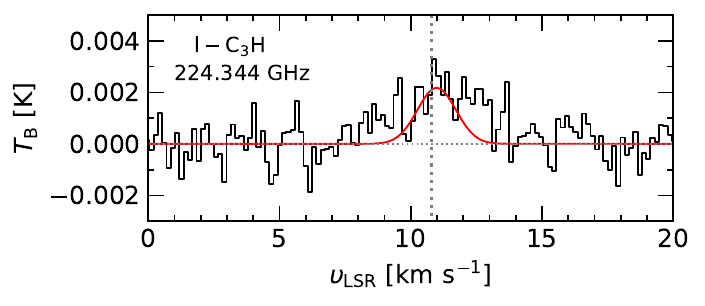}\quad
     \includegraphics[width=0.315\linewidth]{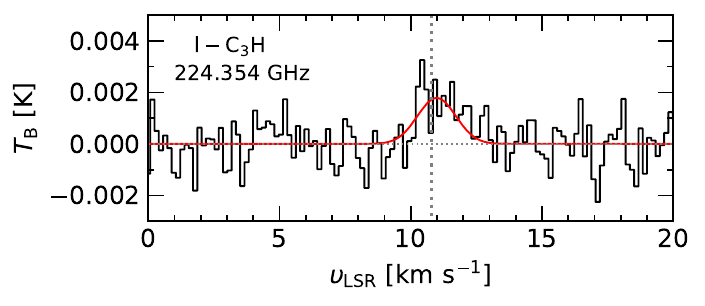} \quad      \includegraphics[width=0.3\linewidth]{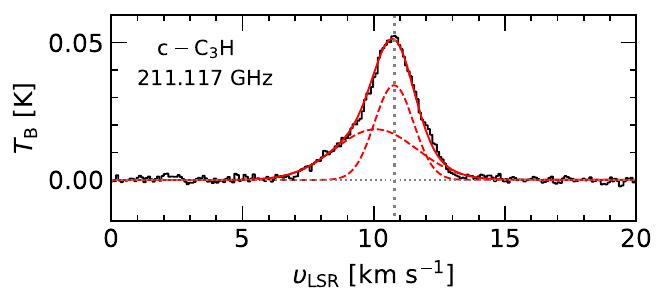} \\
   \includegraphics[width=0.3\linewidth]{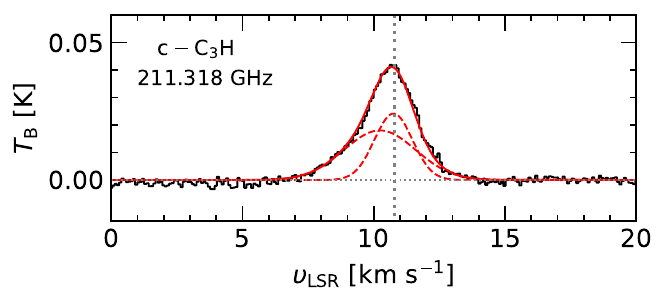} \quad
    \includegraphics[width=0.3\linewidth]{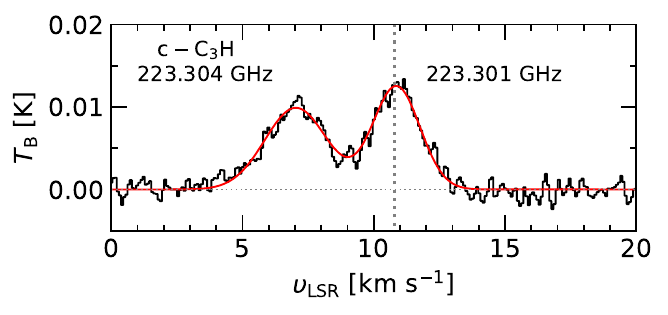} \quad
      \includegraphics[width=0.3\linewidth]{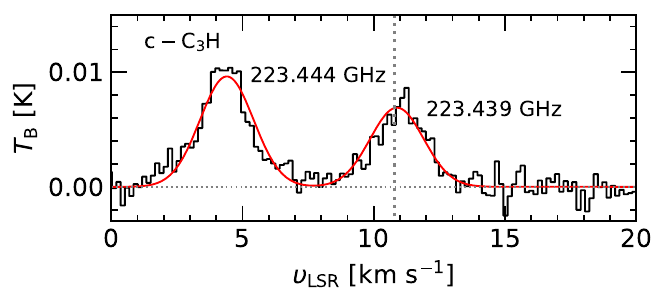}
      \includegraphics[width=0.3\linewidth]{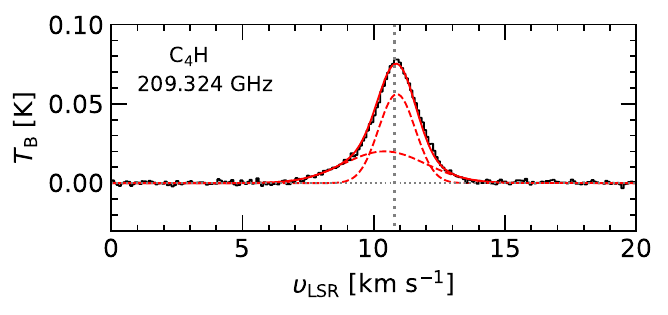}\quad
    \includegraphics[width=0.3\linewidth]{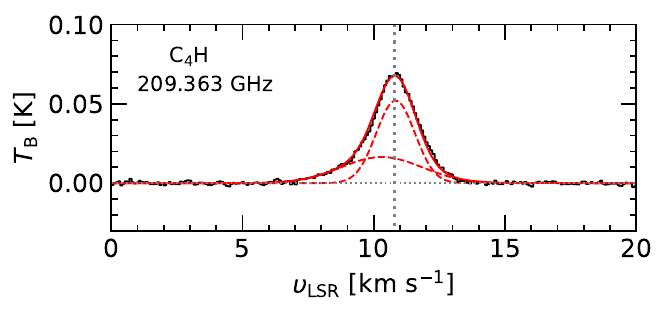}\quad
    \includegraphics[width=0.3\linewidth]{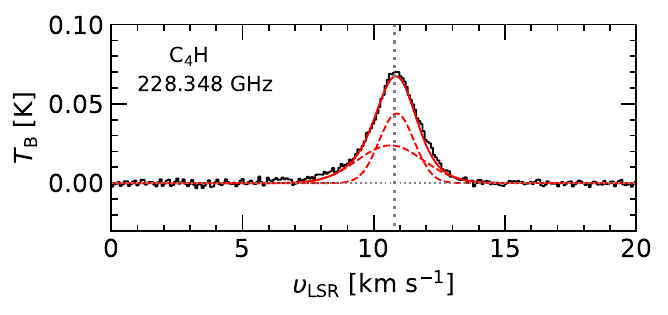}\\
    \includegraphics[width=0.3\linewidth]{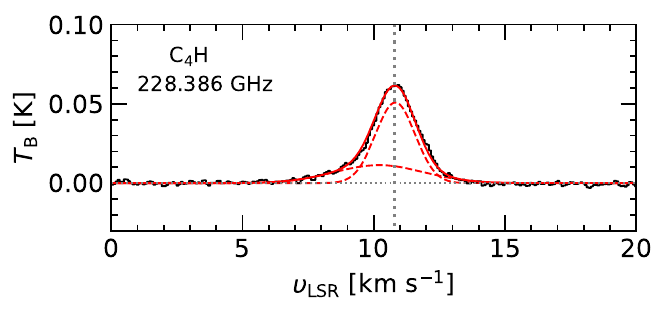} \quad
    \includegraphics[width=0.3\linewidth]{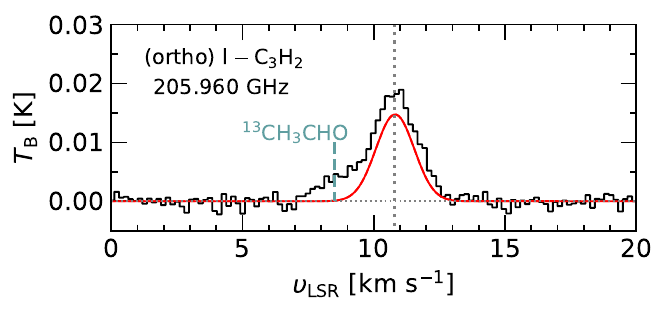}\quad
     \includegraphics[width=0.3\linewidth]{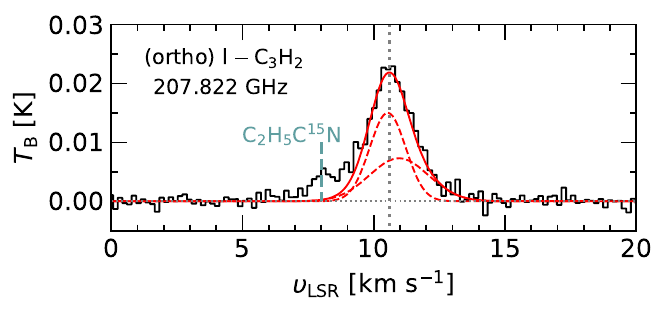} \\      \includegraphics[width=0.3\linewidth]{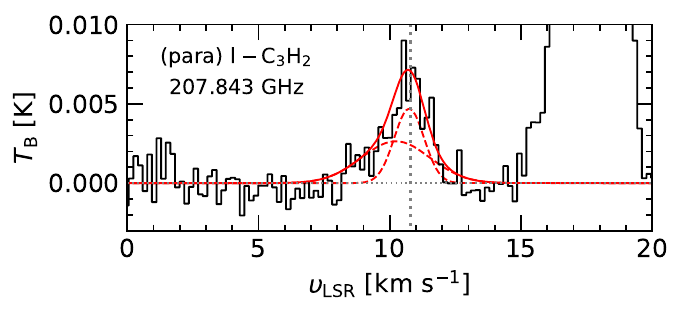} \quad
   \includegraphics[width=0.315\linewidth]{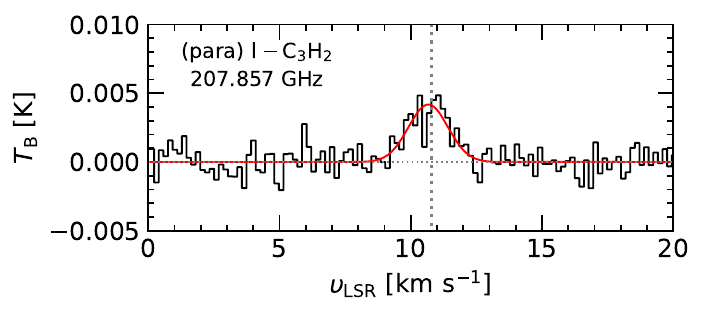} \quad
    \includegraphics[width=0.315\linewidth]{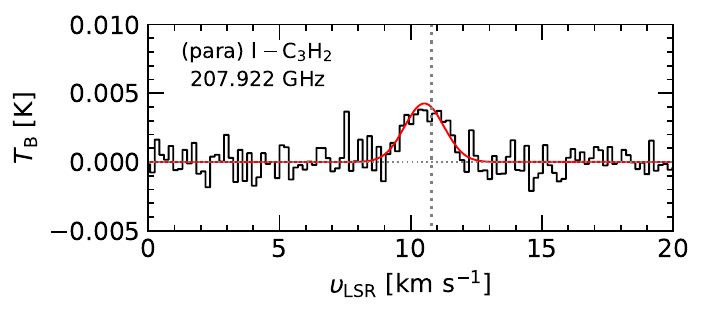} \quad

      \includegraphics[width=0.3\linewidth]{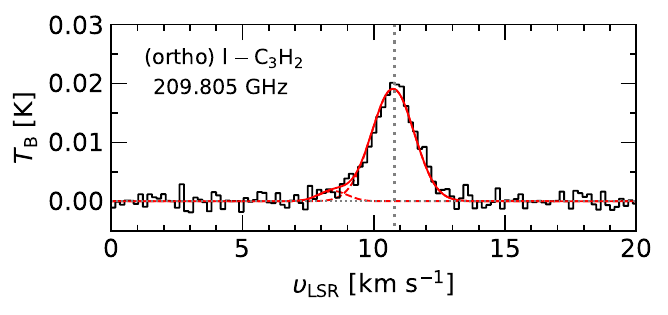}\quad
    \includegraphics[width=0.3\linewidth]{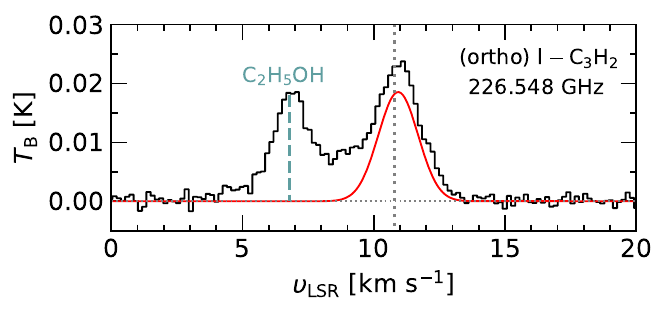}\quad
    \includegraphics[width=0.3\linewidth]{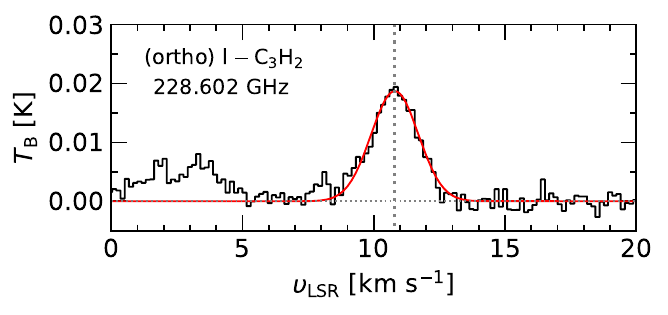}\\
    \includegraphics[width=0.3\linewidth]{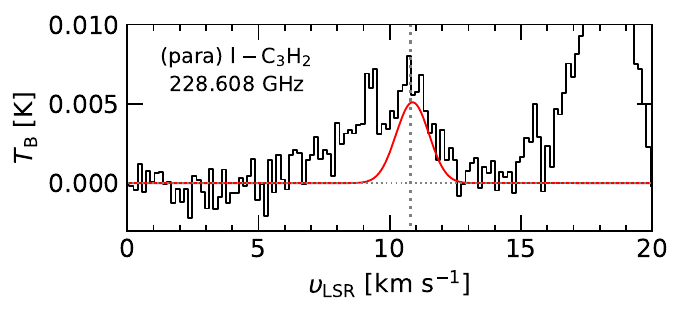} \quad
    \includegraphics[width=0.3\linewidth]{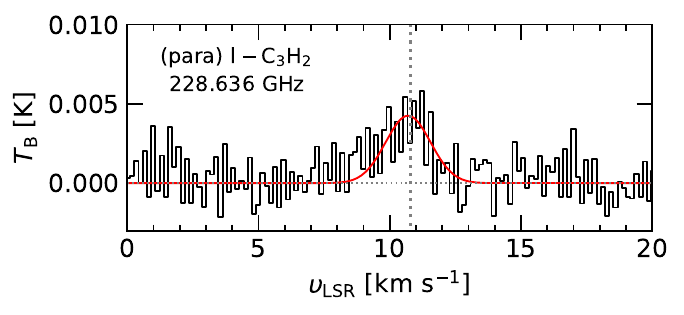}\quad
     \includegraphics[width=0.3\linewidth]{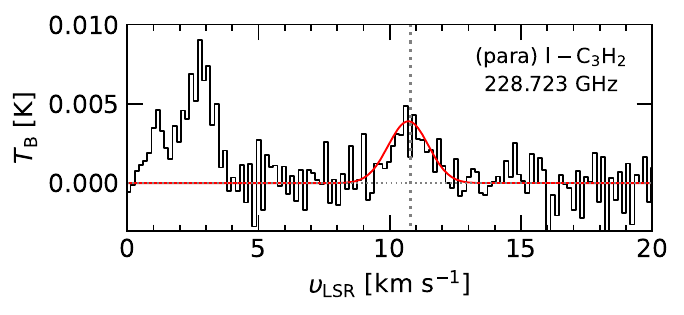} \\      \includegraphics[width=0.3\linewidth]{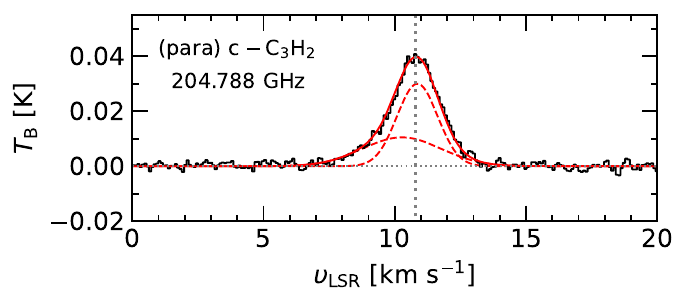} \quad
   \includegraphics[width=0.3\linewidth]{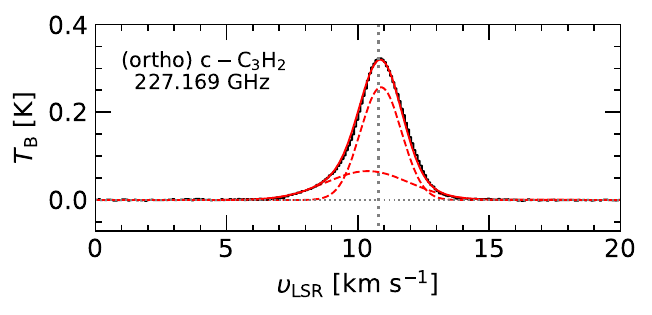} \quad
    
    \caption{Baseline subtracted spectra (black) alongside their multi-component Gaussian fits (dashed red curves) and the total fit (solid red curves), toward detected transitions of C$_3$H$^+$, $l$-/$c$-C$_3$H, C$_4$H, $l$-/$c$-C$_3$H$_2$.}
    \label{fig:spec}
\end{figure*}

\begin{figure*}

\includegraphics[width=0.21\linewidth]{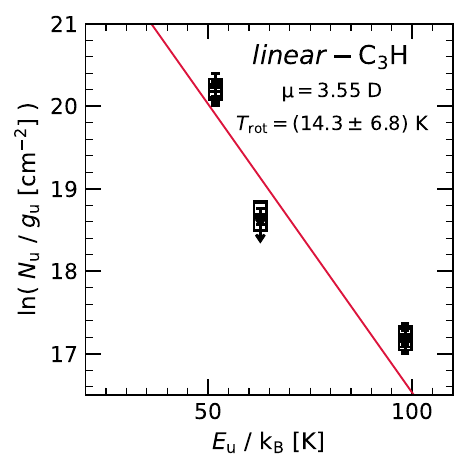} \quad
\includegraphics[width=0.22\linewidth]{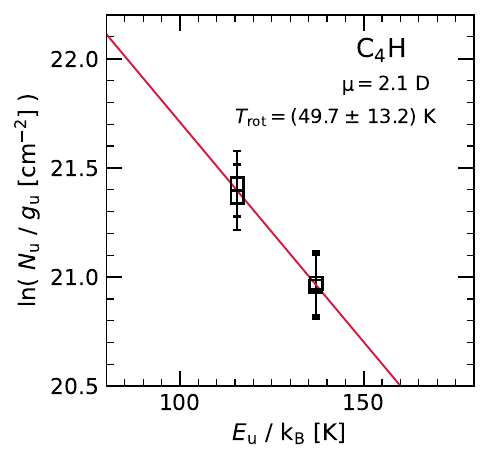}\quad
\includegraphics[width=0.22\linewidth]{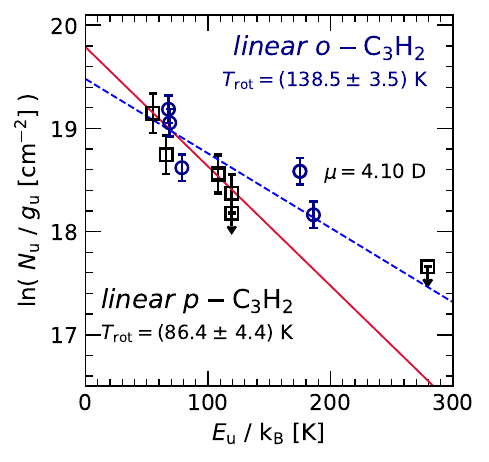}\quad
\includegraphics[width=0.22\linewidth]{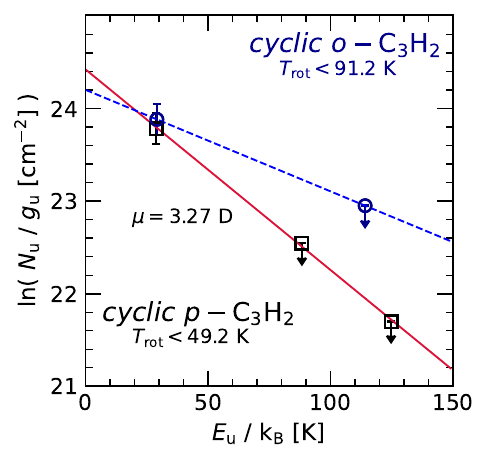}
    \caption{Rotational diagrams of the detected hydrocarbon species in the Orion Bar PDR. The best-fit rotational temperatures, $T_{\rm rot}$ and their associated uncertainties (or upper limits) are indicated for each molecule. The $ortho$ and $para$ nuclear-spin species of $l$- and $c$-C$_3$H$_2$ are shown by blue and black symbols, respectively, with the corresponding fits displayed as solid red and dashed blue lines.}
    \label{fig:rot-diag}
\end{figure*}

The observed spectral line features were fitted with Gaussian profiles in GILDAS-CLASS. Where possible\footnote{This depends on whether multiple transitions of a given species spanning a range of upper-state energies were covered by our observations, enabling rotational temperatures to be constrained via the analysis of population diagrams.}, the Gaussian fitting results were independently compared with models generated using the CLASS-Weeds extension under the assumption of local thermodynamic equilibrium. In such cases, the excitation temperatures adopted in Weeds were constrained using a rotational diagram analysis. The resulting spectral line fits and rotational diagrams, constructed where applicable, are presented in Figs.~\ref{fig:spec} and \ref{fig:rot-diag}. 

Non-detections were not included in the rotational-diagram fits; however, for C$_3$H$_2$, the limited number of detected transitions precluded a meaningful fit, and the quoted rotational temperatures should therefore be regarded as upper limits. Differences in the derived total column density and rotational temperature of C$_4$H with those derived by \citet{Cuadrado2015} can be attributed mainly to the use of the updated dipole moment, recently re-measured in \citet{Oyama2020}. For the other species, larger uncertainties primarily result from insufficient coverage of transitions across a broad range of energy levels, which limits the constraints on the rotational diagrams, as well as from the significant number of non-detections and hence upper limits. Lastly, a rotational diagram analysis was not carried out for $c$-C$_3$H$^+$ because the observations presented in this work did not cover a substantial energy range. 

Rotational temperatures range from $\sim$14 to 138~K. The ortho-to-para ratio obtained from cyclic and linear C$_3$H$_2$ column densities is 3.6$\pm$1.1 and 3.9$\pm$1.6, respectively. These values do not deviate significantly from the equilibrium ortho-to-para ratio of 3.0, expected at high temperatures. The large spread in rotational temperatures is likely because of the limited number of transitions used to constrain the rotational diagram analysis. 
 
\begin{table*}[]
    \caption{Spectroscopic and line-fit parameters of identified hydrocarbons.}
    \centering
    {\small 
    \begin{tabular}{l l c c c c c c c c}
    \hline \hline 
        Species & \multicolumn{1}{c}{Transition} & \multicolumn{1}{c}{Frequency} & $E_{\rm u}$ & $g_{\rm u}$ & $A_{\rm ul}$ &  $\upsilon_{\rm LSR}$ & $\Delta \upsilon$ & $T_{\rm MB}$ & $\int T_{\rm MB} {\rm d}\upsilon $\\
&  & [MHz] & [K] &  & [s$^{-1}$] & [km~s$^{-1}$] & [km~s$^{-1}$] & [mK] & [mK~km~s$^{-1}$] \\
\hline
C$_2$H & $(N,J,F)^{\prime} \rightarrow (N,J,F)^{\prime\prime}$\\
&  ($3,7/2,3 \rightarrow  2,5/2,3$)  &261978.120  &   25.1  &   7 &  $1.96\times10^{-6}$ & 10.41(0.24) &3.41(0.23) & 18.53 & 67.30(4.54) \\
& & & & & & 10.78(0.11) & 1.81(0.30) &  50.96 &  98.23(16.28)     \\
 & ($3,7/2,4 \rightarrow 2,5/2,3$) & 262004.260  &    25.1  &  9 & $5.32\times10^{-5}$  &  10.41(0.24) &3.41(0.23) & 42.65 & 154.88(10.44) \\
& & &  & & & 10.78(0.11) & 1.81(0.30)&  1.76$\times 10^{3}$ & 3.39(0.56)$\times 10^{4}$ \\
 & ($3,7/2,3 \rightarrow  2,5/2,2$) & 262006.482  &    25.1 &   7 & $5.12\times 10^{-5}$  &  10.41(0.24) &3.41(0.23) & 38.36 & 139.30(9.40)\\
& & & & & &  10.78(0.11) & 1.81(0.30) & 1.32$\times 10^{3}$  & 2.54(0.42)$\times10^{4}$\\
   & ($3,5/2,3 \rightarrow  2,3/2,2$)   & 262064.986  &    25.2  &  7 & $4.89\times10^{-5}$ &   10.41(0.24) &3.41(0.23) & 32.21 & 116.97(7.88)\\
& & & & & & 10.78(0.11) & 1.81(0.30)   & 1.26$\times 10^{3}$  & 2.43(1.30)$\times10^{4}$     \\
 & ($3,5/2,2 \rightarrow  2,3/2,1$) &  262067.469   &    25.2  &  5   & $4.47\times10^{-5}$   &   10.41(0.24) &3.41(0.23) & 30.83 & 111.96(7.55)\\
& & & & & & 10.78(0.11) & 1.81(0.30)&  0.82$\times 10^{3}$& 1.58(1.50)$\times10^{4}$ \\
 & ($3,5/2,2 \rightarrow  2,3/2,2$) &   262078.935  &   25.2 &   5 & $6.02\times10^{-6}$   &   10.41(0.24) &3.41(0.23) & 40.45 & 146.89(9.90)\\

& & & & &  & 10.78(0.11) & 1.81(0.30)    & 1.12$\times10^{3}$  & 2.16(0.35)$\times10^{4}$    \\
      & ($3,5/2,2 \rightarrow  2,3/2,2$) & 262208.614   &    25.2  &   7  & $3.96\times10^{-6}$    &   10.41(0.24) &3.41(0.23) & 37.28 & 135.38(9.13)\\
& & & & & & 10.78(0.11) & 1.81(0.30)  & 1.03$\times10^{3}$ &1.98(0.33)$\times10^{4}$ \\
 
  & ($3,5/2,2 \rightarrow  2,5/2,3$) &  262222.586  &    25.2  &  5 & $ 2.37\times10^{-7}$   & -- & -- & -- & -- \\
     & ($3,5/2,3 \rightarrow  2,5/2,2$) &  262236.958   &   25.2  &  7 & $4.04\times10^{-7}$  & -- & -- & -- & -- \\
        & ($3,5/2,2 \rightarrow  2,5/2,2$) &   262250.929   &  25.2   & 5  &  $2.27\times10^{-6}$ &10.41(0.24) &3.41(0.23) & 15.26 & 55.42(3.73)\\

& & & & &  &  10.78(0.11) & 1.81(0.30) &  42.20 & 81.34(13.48)\\

\hline
$l-$C$_3$H$^+$ & $(J^{\prime} \rightarrow J^{\prime\prime})$ \\
& $(10 \rightarrow 9)$ & 224868.2765 & 59.36 &	21 & $5.67\times10^{-4}$ & 10.35(0.07) & 3.81(0.13) & 17.38 & 70.51(5.04)\\
       & & & & & & 10.85(0.01) &  1.78(0.04) &  44.80 & 85.12(5.13)\\
\hline
$l-$C$_3$H & $(J, F)^{\prime} \rightarrow (J, F)^{\prime\prime}$\\
& ($19/2^{-}, 10 \rightarrow 17/2^{+}, 9$) & 207279.369 & 51.8 & 21 & $6.18\times10^{-4}$\rdelim\}{2}{*} & 10.14(0.18)  & 4.12(0.30) & 7.61 & 33.37(5.01)\\
& ($19/2^{-}, 9 \rightarrow 17/2^{+}, 8$) & 207279.779 & 51.8 & 19 & $6.14\times10^{-4}$\,\,\,\, & 10.77(0.04) & 1.77(0.16) & 14.37 & 27.10(4.83) \\
& ($19/2^{-}, 9 \rightarrow 17/2^{+}, 9$) & 207291.938 & 51.8 & 19 & $3.61\times10^{-6}$  & -- & -- & -- & $<6.0$ \\
& ($19/2^{+}, 9 \rightarrow 17/2^{-}, 9$) & 207456.426 & 51.8 & 19 & $3.62\times10^{-6}$  & -- & -- & -- & $<6.0$ \\

& ($19/2^{+}, 10 \rightarrow 17/2^{-}, 9$) & 207459.226 & 51.8 & 21 & $6.19\times10^{-4}$\rdelim\}{2}{*} & 10.17(0.22)  & 3.81(0.33) & 7.66 & 31.11(7.41)\\
& ($19/2^{+}, 9 \rightarrow 17/2^{-}, 8$) & 207459.800 & 51.8 & 19 & $6.16\times10^{-4}$\,\,\,\, & 10.81(0.06) & 1.98(0.21) & 12.81 & 27.07(7.37) \\

& ($21/2,11 \rightarrow 19/2, 10$)\tablefootmark{*} &  224344.903  &  98.3 &  23 &  $7.89\times10^{-4}$\rdelim\}{2}{*} & \multirow{2}{*}{10.99(0.17)} & \multirow{2}{*}{1.72(0.80)} & \multirow{2}{*}{2.18} & \multirow{2}{*}{4.00(1.00)} \\ 
& ($21/2,10 \rightarrow 19/2, 9$) & 224344.948 &      98.3  & 21 & $7.85\times 10^{-4}$\,\,\,\, &\\
& ($21/2,10 \rightarrow 19/2, 10$) & 224348.236   &    98.3  & 21 & $3.66\times10^{-6}$  & -- & -- & -- & $<6.0$ \\
& ($21/2,9 \rightarrow 19/2, 9$) &  224349.849  &    98.3  & 19 & $4.69\times10^{-6}$  & -- & -- & -- & $<6.0$ \\
& ($21/2,10 \rightarrow 19/2, 9$) &  224354.198  &    98.3  & 21 & $7.85\times10^{-4}$\rdelim\}{2}{*} & \multirow{2}{*}{10.99(0.17)} & \multirow{2}{*}{1.72(0.38)} & \multirow{2}{*}{1.78} & \multirow{2}{*}{3.25(0.70)}\\
& ($19/2,9 \rightarrow 17/2, 8$) &  224354.245   &    98.3  & 19 & $7.80\times10^{-4}$\,\,\,\, &\\
& ($25/2,11 \rightarrow 19/2, 10$) &  224386.532 &  98.3  & 23 & $9.27\times10^{-9}$  & -- & -- & -- & $<6.0$ \\

& $(21/2^+,11 \rightarrow 19/2^{-},10)$ & 229213.695   &   62.8  & 23 &  $6.45\times10^{-4}$\rdelim\}{2}{*}  & \multirow{2}{*}{--} & \multirow{2}{*}{--} & \multirow{2}{*}{--} & \multirow{2}{*}{$<6.0$} \\
& $(21/2^+,10 \rightarrow 19/2^{-},9)$ &229214.063  &  62.8  & 21 &  $6.42\times10^{-4}$\,\,\,\, & \\
& $(21/2^+,10 \rightarrow 19/2^{-},10)$ & 229226.596  &   62.8  & 21 & $3.07\times 10^{-6}$  & -- & -- & -- & $<6.0$ \\
& $(21/2^-,10 \rightarrow 19/2^{+},10)$ &  229430.543 &  62.8  & 21 & $3.05\times10^{-6}$  & -- & -- & -- & $<6.0$ \\
& $(21/2^-,11 \rightarrow 19/2^{+},10)$ &  229432.802 &  62.8  & 23 & $6.41\times10^{-4}$\rdelim\}{2}{*} & \multirow{2}{*}{--} & \multirow{2}{*}{--} & \multirow{2}{*}{--} & \multirow{2}{*}{$<6.0$} \\
& $(21/2^-,10 \rightarrow 19/2^{+},9)$ &  229433.336 &  62.8  & 21 & $6.38\times10^{-4}$\,\,\,\, & \\

$c-$C$_3$H & $(N_{K_{\rm a}K_{\rm c}}, J, F)^{\prime} \rightarrow (N_{K_{\rm a}K_{\rm c}}, J, F)^{\prime\prime}$\\
& $(5_{1,5}, 11/2, 6) \rightarrow (4_{1,4}, 9/2, 5)$  & 211117.576 & 29.2 & 13 & $2.74\times10^{-4}$\rdelim\}{2}{*} & 10.07(0.06) & 3.70(0.08) & 18.46 & 72.58(3.70)\\
& $(5_{1,5}, 11/2, 5) \rightarrow (4_{1,4}, 9/2, 4)$  & 211117.834 & 29.2 & 11 & $2.68\times10^{-4}$\,\,\,\, & 10.77(0.01) & 1.75(0.04) & 34.47 & 64.10(3.65)\\

& $(5_{1,5}, 11/2, 5) \rightarrow (4_{1,4}, 9/2, 5)$ & 211137.462  &  29.2  & 11  & $5.11\times10^{-6}$  & -- & -- & -- & $<6.0$ \\
& $(5_{1,5}, 9/2, 4) \rightarrow (4_{1,4}, 7/2, 4)$& 211298.853  &   29.2  &  9 & $7.23\times10^{-6}$ & -- & -- & -- & $<6.0$ \\
 
& $(5_{1,5}, 9/2, 4) \rightarrow (4_{1,4}, 7/2, 3)$  & 211318.450 & 29.2 & 9 & $2.61\times10^{-4}$\rdelim\}{2}{*} & 10.23(0.06) & 3.21(0.09) & 18.04 & 61.64(4.86)  \\
& $(5_{1,5}, 9/2, 5) \rightarrow (4_{1,4}, 7/2, 4)$  & 211318.796 & 29.2 & 11 & $2.68\times10^{-4}$\,\,\,\, & 10.75(0.02) & 1.70(0.07) & 24.15 & 43.82(4.85)\\
& $(5_{1,5}, 9/2, 4) \rightarrow (4_{1,4}, 9/2, 4)$ & 211957.280  &  29.2 &    9  & $6.20\times 10^{-6}$ & -- & -- & -- & $<6.0$ \\
& $(5_{1,5}, 9/2, 4) \rightarrow (4_{1,4}, 9/2, 5)$ & 211976.908  &  29.2  &  9 & $1.36\times 10^{-7}$ & -- & -- & -- & $<6.0$ \\
& $(5_{1,5}, 9/2, 5) \rightarrow (4_{1,4}, 9/2, 4)$ & 211977.223  &  29.2  &  11 & $7.21\times 10^{-8}$  & -- & -- & -- & $<6.0$ \\
& $(5_{1,5}, 9/2, 5) \rightarrow (4_{1,4}, 9/2, 5)$& 211996.851 &  29.2  &  11  & $5.88\times10^{-6}$ & -- & -- & -- & $<6.0$ \\

& $(4_{3,2}, 7/2, 3) \rightarrow (3_{3,1}, 7/2, 3)$ & 223167.731  &  30.9 &   7  & $4.56\times10^{-6}$   & -- & -- & -- & $<6.0$ \\
& $(4_{3,2}, 7/2, 4) \rightarrow (3_{3,1}, 7/2, 3)$ & 223175.963  & 30.9  &  9 & $6.85\times10^{-8}$ & -- & -- & -- & $<6.0$ \\
& $(4_{3,2}, 7/2, 3) \rightarrow (3_{3,1}, 7/2, 4)$ & 223180.147  & 30.9  &  7 & $1.96\times10^{-7}$  & -- & -- & -- & $<6.0$ \\
& $(4_{3,2}, 7/2, 4) \rightarrow (3_{3,1}, 7/2, 4)$ & 223188.379  &  30.9  &  9  & $6.39\times10^{-6}$ & -- & -- & -- & $<6.0$\\\

& $(4_{3,2}, 9/2, 5) \rightarrow (3_{3,1}, 7/2, 4)$ & 223301.350 &  30.9  &  11  & $1.53\times10^{-4}$  & 10.86(0.02) & 2.11(0.05) & 12.50 & 28.11(0.06)\\
& $(4_{3,2}, 9/2, 4) \rightarrow (3_{3,1}, 7/2, 3)$ & 223304.289 &  30.9 &    9  & $1.50\times10^{-4}$ & 10.93(0.04) & 2.78(0.11) & 9.90 & 29.36(0.08)    \\
& $(4_{3,2}, 9/2, 4) \rightarrow (3_{3,1}, 7/2, 4)$& 223316.704  &  30.9  &   9  & $3.20\times10^{-6}$ & -- & -- & -- & $<6.0$\\

& $(4_{3,2}, 7/2, 3) \rightarrow (3_{3,1}, 5/2, 3)$ &  223436.471  &  30.9  &   7 & $7.77\times10^{-6}$   & -- & -- & -- & $<6.0$ \\
& $(4_{3,2}, 7/2, 3) \rightarrow (3_{3,1}, 5/2, 2)$ & 223439.723   & 30.9  &  7  & $1.41\times10^{-4}$  & 10.88(0.04) & 2.40(0.11) & 6.91 & 17.63(0.07)\\
& $(4_{3,2}, 7/2, 4) \rightarrow (3_{3,1}, 5/2, 3)$ & 223444.702   & 30.9  &  9  & $1.47\times10^{-4}$ & 10.87(0.03) & 2.44(0.09) & 9.64 & 25.06(0.07)\\
& $(4_{3,2}, 9/2, 4) \rightarrow (3_{3,1}, 5/2, 3)$ & 223573.028 & 30.9  &  9  & $4.08\times 10^{-7}$  & -- & -- & -- & $<6.0$ \\

\hline

   \end{tabular}

    \label{tab:other_lines1}
    }
\end{table*}

\addtocounter{table}{-1} 
\begin{table*}[]
    \caption{Continued.}
    {\small 
    \centering
    \begin{tabular}{l l l c c c c c c c }
    \hline \hline 
        Species & \multicolumn{1}{c}{Transition} & \multicolumn{1}{c}{Frequency} & $E_{\rm u}$ & $g_{\rm u}$ & $A_{\rm ul}$ &  $\upsilon_{\rm LSR}$ & $\Delta \upsilon$ & $T_{\rm MB}$ & $\int T_{\rm MB} {\rm d}\upsilon $\\
&  & [MHz] & [K] &  & [s$^{-1}$] & [km~s$^{-1}$] & [km~s$^{-1}$] & [mK] & [mK~km~s$^{-1}$] \\
        \hline

       C$_4$H & $(N, J, F)^{\prime} \rightarrow (N, J, F)^{\prime\prime}$ & \\
       & $(22, 45/2, 22) \rightarrow (21, 43/2, 21)$ & 209324.9164 &  115.53 & 45 & $2.30 \times 10^{-4}$\rdelim\}{2}{*} & 10.41(0.05) & 3.70(0.08) & 20.05 & 80.54(4.17)  \\
       & $(22, 45/2, 23) \rightarrow (21, 43/2, 22)$ & 209324.9179 &  115.53 & 47 & $2.30 \times 10^{-4}$\,\,\,\, & 10.88(0.07) & 1.65(0.03) & 56.11 & 98.45(4.13)  \\
       & $(22, 43/2, 21) \rightarrow (21, 41/2, 20)$ & 209363.2935 &  115.57 & 43 & $2.30 \times 10^{-4}$\rdelim\}{2}{*} &  10.30(0.06) & 3.62(0.10) & 16.5 & 64.56(4.30) \\
       & $(22, 43/2, 22) \rightarrow (21, 41/2, 21)$ & 209363.2944 &  115.57 & 45 & $2.30 \times 10^{-4}$\,\,\,\, & 10.83(0.10)  &    1.73(0.03) & 52.00 &  95.82(4.34)\\
& $(24, 49/2, 24) \rightarrow (23, 47/2, 23)$ & 228348.5990 & 136.99 & 49 & $2.99\times 10^{-4}$\rdelim\}{2}{*} & 10.63(0.04) & 2.95(0.10) & 23.83 & 74.73(5.42)\\
& $(24, 49/2, 25) \rightarrow (23, 47/2, 24)$ & 228348.6003 & 136.99 & 51 & $2.99 \times 10^{-4}$\,\,\,\, & 10.87(0.01) & 1.57(0.04) & 43.93 & 73.72(5.48)\\
& $(24, 47/2, 23) \rightarrow (23, 45/2, 22)$ & 228386.9394 & 137.03 & 47 & $2.99\times 10^{-4}$\rdelim\}{2}{*} &  10.20(0.15) & 4.02(0.26) & 11.38 & 48.75(6.24) \\
& $(24, 47/2, 24) \rightarrow (23, 45/2, 23)$ & 228386.9402 & 137.03 & 49 & $2.99\times 10^{-4}$\,\,\,\, &    10.82(0.01) & 1.77(0.05) & 50.83 & 95.95(6.20)\\
\hline

$l-$C$_3$H$_2$ & $(J, F)^{\prime} \rightarrow (J, F)^{\prime\prime}$\\

& $10_{1,10} \rightarrow 9_{1,9}$, \,\,  \textit{ortho}\tablefootmark{*} & 205960.125 & 67.7 &  63 & $8.06\times 10^{-4}$  & 10.81(0.03)&      1.71(0.11)  & 14.74 & 26.91(3.94)\\

& $10_{3,8} \rightarrow 9_{3, 7}$, \,\, \textit{ortho}\tablefootmark{*} & 207822.296  &   175.0 &   63 & $7.61\times10^{-4}$\rdelim\}{2}{*} & 10.96(0.46) &2.64(0.46) &  7.30 & 20.50(12.67) \\
& $10_{3,7} \rightarrow 9_{3, 6}$, \,\,  \textit{ortho}\tablefootmark{*} & 207822.494  &  175.0  & 63 & $7.61\times10^{-4}$\,\,\,\, & 10.53(0.06) &    1.61(0.18) & 15.02 & 25.71(10.80)\\
& $10_{0,10} \rightarrow 9_{0,9}$, \,\,  \textit{para}  & 207843.289 &  54.9  & 21 & $8.37\times10^{-4}$  &  10.23(0.31) &    3.08(0.55) & 2.65 & 8.19(3.25) \\
& & & & & &  10.74(0.06)  &   1.34(0.31) &  4.68 &  6.66(3.12) \\

& $10_{2,9} \rightarrow 9_{2,8}$, \,\, \textit{para} & 207857.566  & 108.3  & 21  & $8.03\times10^{-4}$ &  10.67(0.06) &  1.77(0.13) & 4.18 & 7.85(0.51)\\

& $10_{2,8} \rightarrow 9_{2,7}$, \,\, \textit{para} & 207922.778  &  108.3 &  21  & $8.04\times10^{-4}$ & 10.52(0.07)  &   1.76(0.15) &  4.26 &   7.97(0.60)  \\

 & $10_{1,9} \rightarrow 9_{1,8}$, \,\, \textit{ortho} & 209805.427 & 68.7 &  63 & $8.52\times10^{-4}$  & 8.45(0.16) &  1.30(0.30) & 1.76 &  2.42(0.60)\\
 & & &  & &  & 10.73 (0.02) & 1.91(0.05)  & 19.10 &   38.76(0.75) \\

 & $11_{1,11} \rightarrow 10_{1, 10}$, \,\, \textit{ortho}\tablefootmark{*} & 226548.575   &   78.6  & 69&  $1.08\times 10^{-3}$  &  10.93(0.02)  &   1.78(0.05) &  18.54 & 35.18(1.71) \\

 & $11_{4, 8} \rightarrow 10_{4, 7}$, \,\, \textit{para}  & 228508.904 & 279.3  & 23 & $9.69\times10^{-4}$\rdelim\}{2}{*}&  \multirow{2}{*}{10.80(0.10)} & \multirow{2}{*}{2.36(0.22)} & \multirow{2}{*}{3.91} & \multirow{2}{*}{9.82(0.81)}\\
  & $11_{4, 7} \rightarrow 10_{4, 6}$, \,\, \textit{para}  & 228508.905 
  & 279.3  & 23 &  $9.69\times10^{-4}$\,\,\,\, &      \\

  & $11_{3, 9} \rightarrow 10_{3, 8}$, \,\, \textit{ortho} &  228602.691  & 186.0 & 69 &  $1.03\times10^{-3}$\rdelim\}{2}{*} & \multirow{2}{*}{10.70(0.03)} &  \multirow{2}{*}{2.12(0.07)} & \multirow{2}{*}{18.69} & \multirow{2}{*}{42.26(1.22)}   \\
 & $11_{3, 8} \rightarrow 10_{3, 7}$, \,\, \textit{ortho} & 228602.691 & 186.0 & 69 & $1.03\times10^{-3}$\,\,\,\, &   \\

 & $11_{0, 11} \rightarrow 10_{0, 10}$, \,\, \textit{para}\tablefootmark{*} & 228608.345 &    65.8  & 23 & $1.12\times10^{-3}$ &   10.87(0.05) &     2.03(0.18) & 5.81 &  13.00(0.72)\\

& $11_{2,10} \rightarrow 10_{2,9}$, \,\, \textit{para} &     228636.938  &   119.3   & 23 & $1.08\times10^{-3}$ &   10.70(0.10) & 2.01(0.27) &  4.25 &  9.10(0.93)    \\

& $11_{2,9} \rightarrow 10_{2,8}$, \,\, \textit{para}&  228723.873   &   119.3 &  23 &  $1.08\times10^{-3}$ &     10.71(0.14) &    1.80(0.35)  & 3.90 & 7.49(1.23) \\

$c-$C$_3$H$_2$ & \\

& $4_{2,2} \rightarrow 3_{3,1}$, \,\, \textit{para} &   204788.926  & 28.8 &  9 &   $1.24\times 10^{-4}$ & 10.27(0.14) & 3.52(0.23)  & 10.51 & 39.38(6.61) \\
& & &  & & & 10.86(0.02)  & 1.82(0.08) & 30.00 & 58.00(6.66)\\

& $4_{3,2} \rightarrow 3_{2,1}$, \,\, \textit{ortho} & 227169.138   & 29.1 &  27 & $3.11\times10^{-4}$  & 10.37(0.01) &  3.72(0.02) & 66.01 & 261.20(0.40)\\
& & & & & &   10.88(0.01)   &  1.84(0.04) & 257.17 &  503.13(0.13) \\

& $7_{6,1} \rightarrow 7_{3,4}$, \,\, \textit{para} & 230686.032  &    88.3 &   45 & $5.64\times10^{-6}$ & -- & -- & -- & $<6.0$\\
& $8_{7,1} \rightarrow 8_{4,4}$, \,\, \textit{ortho} & 232194.879   & 114.1  &  17 & $1.00\times10^{-5}$ & -- & -- & -- & $<6.0$\\
& $9_{4,5} \rightarrow 8_{7,2}$, \,\, \textit{para} &  234198.724  &  124.7  & 57 & $1.05\times10^{-6}$ & -- & -- & -- & $<6.0$\\
         \hline
    \end{tabular}
 \tablefoot{Integrated intensity of the line features detected at the source systemic velocity with uncertainties presented in parentheses. When not detected we report 3~$\sigma$ upper limits of the integrated intensities, computed across typical line width of 1.8~km~s$^{-1}$. \tablefoottext{*}{Transitions contaminated by other species}.}
    \label{tab:other_lines}
    \tablebib{$l$-C$_3$H$^+$: \citet{Bruenken2014}, $l$-/$c$-C$_3$H: \citet{Gottlieb1983, Yamamoto1990}, C$_4$H: \citet{Gottlieb1983}, $l$-/$c$-C$_3$H$_2$: \citet{Thaddeus1985, V1990ApJ...364L..53V, Lovas1992, Spezzano2012}.}
    }
\end{table*}

\end{appendix}

\end{document}